\documentclass[12pt, preprint]{aastex}

\newcommand{\myemail}{cuk@astro.cornell.edu}

\slugcomment{To appear in Astron. J.}

\shorttitle{Irregular Satellites}
\shortauthors{\' Cuk and Burns}

\begin{document}

\title{On the Secular Behavior of Irregular Satellites}

\author{Matija \' Cuk and Joseph A. Burns\altaffilmark{1}}
\affil{Astronomy Department, Cornell University,
    Ithaca, NY 14853}
\email{\myemail}
\altaffiltext{1}{Department of Theoretical \& Applied Mechanics, Cornell University, Ithaca, NY 14853}

\begin{abstract}
Although analytical studies on the secular motion of the irregular satellites have been published recently \citep{kin99, yok03}, these theories have not yet been satisfactorily reconciled with the results of direct numerical integrations \citep{car02, nes03}. These discrepancies occur because in secular theories the disturbing function is generally averaged over the Sun's orbital motion, whereas instead one should take into account some periodic terms, most notably the so-called ``evection'', which can be large for distant, slow-moving satellites. This problem is identical to that initially encountered by Newton and other historical researchers when studying the Moon's motion. Here we demonstrate that the evection and other terms from lunar theory can be incorporated into the more modern Kozai formalism, and that our synthetic approach produces much better agreement with results from symplectic integrations. Using this method, we plot the locations of secular resonances in the orbital-element space inhabited by the irregular satellites. Our model is found to predict correctly those satellites that are resonant or near-resonant. 

We also analyze the octupole term in the disturbing function \citep{yok03} to determine the strengths of resonant-locking for satellites whose longitudes of pericenter are librating. By independently integrating these satellites' nominal orbits using a symplectic integrator, we show that the strength of this resonance can be successfully obtained from simple analytical arguments. 

We note that the distribution of irregular satellite clusters in the space of proper orbital elements appears to be non-random. We find that the large majority of irregular-satellite groups cluster close to the secular resonances, with several objects (Pasiphae, Sinope, Siarnaq, formerly S/2000~S3, and Stephano, formerly S/1999~U2) having practically stationary pericenters. After proposing the name ``Main Sequence'' to describe this grouping, we point out that none of the largest satellites (those with radii $R >$ 100 km) belong to this class. Finally, we argue that this dichotomy implies that the smaller near-resonant satellites might have been captured differently than the largest irregulars. 
\end{abstract}

\keywords{celestial mechanics --- planets and satellites: general --- 
planets and satellites: individual(\objectname{Pasiphae},
\object{Siarnaq}, \objectname{Sinope},
\object{Stephano})}

\section{Introduction}

Irregular satellites are usually considered to be those that orbit the giant planets far beyond the major moons, often with quite eccentric and inclined orbits. The number of known irregular satellites has grown rapidly \citep{gla98, gla00, gla01, she03, hol04} at the turn of the 21$^{\rm st}$ century due to advances in CCDs and computing, coupled with the availability of numerous large telescopes. The new objects tend to group into clusters \citep{gla01} that are, at least in the case of the Jovian satellites, centered on large objects that have been known for most of the last century. While modern {\it observers} have improved upon the efforts of their predecessors (and sometimes spectacularly so, with the year 2000 re-discovery of Themisto, that was lost since 1975), the {\it theoreticians} have not been similarly active in bettering analytical models, instead generally preferring numerical simulations for determining histories \citep{car02, car04, nes03, ast03}. Thus, for the most part, papers from two generations ago \citep{bro30, bro37} which addressed Pasiphae's motion analytically, have been almost forgotten, even though they treat this problem in more detail than any of the modern work. Of all the contemporary researchers, only \citet{saha93}, while concentrating on a numerical approach, have correctly attributed Pasiphae's secular\footnote{The term ``secular'' refers to averaged or non-periodic behavior, and more specifically to perturbations which occur on a precessional timescale rather than on one comparable to an orbital period.} resonance to mutual cancellation between various terms that appear in the expression for the precession of apsides that was obtained in lunar theory. None of the more recent papers \citep{kin99, car02, yok03, nes03} reference Brown's work, nor do they mention evection and other corrections to secular behavior that arise when the disturbing function is appropriately averaged.

The difficulty in describing the secular behavior of a distant satellite is most famously illustrated by Newton's inability to explain the precession of the lunar apsides. The precession rate of an ecliptic, near-circular satellite orbit under the solar influence, to the first order, is given by

\begin{equation}
\label{newton}
{\dot \varpi}={3 \over 4} \ {n_p^2 \over n_s},
\end{equation}

\noindent where $\varpi$ is the satellite's longitude of pericenter, $n_p$ is the planet's mean motion, $n_s$ is that of the satellite and the dot signifies differentiation with respect to time \citep{dan92}. For Earth's Moon, this formula gives an apsidal {\it precession} period of about 18 yrs, identical to that of the nodal {\it regression}. However, the observed precession of the lunar apsides is twice as fast. Newton himself was frustrated by this discrepancy, and considered it to be a significant failing for his theory of gravitation. Clairaut, in 1847, realized that this large error in precession rate arises from evection terms which remain after averaging over the orbital motion has been completed \citep{baum97}.

Lunar evection, known to Hipparchus, and nicely treated by \citet{bro61}, is, in essence, a periodic perturbation to a satellite's orbit that depends on the angle between the satellite's line of apsides and the Sun's instantaneous position. The relative strength of this perturbation is determined by $m=n_p/n_s$, the ratio of the mean motions of the planet and the satellite (see Eq. 1). For Earth's Moon, $m$ is about 1/13, while it can be as large as 1/5 for some retrograde Jovian satellites. In the next section, we show that, for an $m$ of only 0.1, evection-induced precession of the apsides surpasses in magnitude that arising from purely secular terms.

When the Space Age started nearly fifty years ago, the focus for the hierarchical three-body problem (Sun, planet, satellite) veered away from the motions of the Moon and other natural satellites to the orbits of artificial satellites; more recently, planets in binary systems \citep{inn97, hol99} were studied. This shift has redirected the emphasis from a very accurate description of orbits that are close to being circular and equatorial,  to a more qualitative treatment of orbits having arbitrary inclinations and eccentricities. Generally, numerical integrators have been employed whenever the celestial bodies' positions need to be accurately predicted. 

Analytical models of the hierarchical three-body problem usually average the leading term (in $a_{s}/a_{p} \sim m^{2/3}$) of the disturbing function over the orbital motions of both perturber and the perturbee, and obtain the differential equations for the evolutions of inclination $i$, eccentricity $e$, argument of pericenter $\omega$ and longitude of ascending node $\Omega$ for arbitrary eccentricity and inclination (Innanen et al. 1997, corrected by Carruba et al. 2003):

\begin{equation}
\label{kozaii}
{di \over d\tau} = - {15 \over 8} e^2 {\sin}(2 \omega) \ {\sin} i \ {\cos} i/ \sqrt {1-e^2} , 
\end{equation} 

\begin{equation}
\label{kozaie}
{de \over d\tau} = {15 \over 8} e \ \sqrt {1-e^2} {\sin}(2 \omega) {\sin}^2 i ,
\end{equation}

\begin{equation}
\label{kozaiw}
{d\omega \over d\tau} = {3 \over 4} \{ 2 (1-e^2)+5 \ {\sin}^2 \omega \lbrack e^2-{\sin}^2 i \rbrack \} / \sqrt {1-e^2} ,
\end{equation}

and

\begin{equation}
\label{kozaio}
{d\Omega \over d\tau} = - {{\rm cos} i \over 4} \lbrack 3+12e^2-15e^2 \ {\cos}^2 \omega  \rbrack / \sqrt {1-e^2} ,
\end{equation}

\noindent where time has been non-dimensionalized by introducing $\tau=(1-e_p^2)^{-3/2} m n_p t$, with $e_p$ being the planet's heliocentric eccentricity. The above equations can either be integrated numerically \citep{car02, nes03} or their solutions can be expressed in terms of elliptical integrals \citep{kin99}. 

A satellite's secular behavior obtained from this low-order theory can be divided into cases when the argument of pericenter {\it circulates} or {\it librates}. According to this simplified theory, for all satellites with inclinations below 39.2$^{\circ}$ (or greater than 140.8$^{\circ}$), $\omega$ circulates from 0 to $2 \pi$, while for orbits with higher inclinations ($39.2^\circ < i < 140.8^\circ$) $\omega$ either circulates or librates around 90$^{\circ}$ or 270$^{\circ}$. The last case is what is usually referred to as the \underline{``Kozai resonance''}, and was among the most important discoveries of the ``new celestial mechanics'' for the hierarchical three-body problem \citep{koz62}.  The first natural satellites observed to be in the Kozai resonance were Saturn's Kiviuq and Ijiraq \citep{vas01, cuk02a, nes03, yok03}; now about a half-dozen are known.

However, as we have gradually come to realize, the Kozai theory, as developed so far, is seriously limited. We have already mentioned that, for small eccentricity and inclination, the formula for the precession of $\varpi$ (defined as $\Omega+\omega$ for prograde bodies) reduces to Eq. \ref{newton}, and gives an error of 50\% for the Moon. Furthermore,  the Kozai model suggests that prograde and retrograde satellites should have symmetric secular behavior: yet, while the Moon's apsides precess faster than they ``should'', retrograde satellites like Jupiter's Pasiphae and Sinope show very slow precessions of their apsides, and it is this that enables a secular resonance to take effect \citep{whi93, saha93}. Even though the strength of the locking between the planet's and the satellite's lines of apsides depends on the octupole term in the disturbing function (Yokoyama et al. 2003; Sec. 4 below), the resonance first requires that the leading terms give ${\dot \varpi} \simeq 0$, which is impossible according to Eq. \ref{newton} and becomes possible only for a specific inclination ($\simeq 36^{\circ}$ for $e \simeq 0$) in Eqs. \ref{kozaii} to \ref{kozaio}. It is historically interesting to note that Yoshihide Kozai himself, in the very last paragraph of his landmark paper \citep{koz62}, mentioned that his method cannot be applied to the most distant satellites, due to problems with averaging over the solar true longitude.

Many researchers have worked on the lunar theory since Newton and Clairaut \citep{tis94, bro96, bro61}, and -- for this case of low $e$ and $i$ -- they have obtained an expression for ${\dot \varpi}$ that is a high-order polynomial in $m$ \citep{tis94}:

\begin{equation}
\label{moon}
{\dot \varpi}= n_{p} {\Bigl(\ {3 \over 4} m + {225 \over 32} m^2 + {4071 \over 128} m^3 + {265493 \over 2048} m^4 ...\Bigr)\ .}
\end{equation}

\noindent This expression makes clear that there may be (negative) $m$'s for which $\dot \varpi =0$. Saha \& Tremaine (1993, their Fig. 1) recognized Eq. 6's relevance in determining the secular resonances of Pasiphae and Sinope. The second term, which invariably produces prograde motion, derives solely from evection, as we will show next. Although evection is the largest correction to the purely secular motion, the convergence of the series is slow (because higher-power terms in $m$ have larger coefficients as seen in Eq. \ref{moon}) and usually terms with powers of $m$ as high as 10 are listed. Higher-order terms come from different sources (mostly from periodic terms that do not completely average out) and their calculation requires a quite specialized theory.

Unfortunately, the classical approaches used to derive Eq. \ref{moon} are not practical for irregular-satellite applications. All lunar theories assume small eccentricity and inclination, and are always constructed in specific variables which are significantly removed from the usual orbital elements. The only instances when the ``old celestial mechanics'' was applied to the study of an irregular satellite's orbit were a pair of papers by E.W. Brown on Pasiphae (1930, Brown \& Brouwer 1937). In our opinion, Brown's approach is remarkably difficult, yet yields only approximate results.

The main goal of this paper is to incorporate evection and other terms from Eq. \ref{moon} into Kozai's set of equations. To do this, in the next chapter we explicitly derive the dependence of the evection term on eccentricity and inclination, and then argue how other, higher-order terms should behave with inclination. In the subsequent chapters we compare our results to those from direct numerical integrations, and discuss the possible significance of secular resonances for the origin of the irregular satellites.

\section{Development of the Disturbing Function}

As usual, we consider the planet to be the central body and the Sun to be an exterior perturber. In that case, the disturbing function describing the solar influence on the satellite is \citep{md99}:

\begin{equation}
\label{legendre}
R = {\mu ' \over r '} \sum_{l=2}^{\infty} \Bigl( {r \over r '} \Bigr)^l P_l ({\cos} \psi),
\end{equation}

\noindent where $\mu '$ is solar mass times the gravitational constant, $r$ and $r '$ are the distances of the satellite and the Sun, respectively, from the planet, while $\psi$ is the angle between their position vectors. $P_l ({\cos} \psi)$ are Legendre polynomials of order $l$ with the argument $\cos \psi$. Since the ratio $r / r'$ is small, of $O(m^{2/3})$, for most applications only the leading ($l=2$, or quadrupole) term in the disturbing function is taken. The next term ($l=3$, or octupole), which is important only within secular resonances, will be discussed later. The quadrupole term is

\begin{equation}
\label{quadrupole}
R = {\mu '}  \Bigl( {r^2 \over {r '}^3} \Bigr) \ {1 \over 2} (3 {\cos}^2 \psi - 1).
\end{equation}

From spherical trigonometry, we can expand the angle $\psi$ as:
\begin{equation}
\label{psi}
\cos \psi=\cos(\lambda ' - \Omega) \cos (\omega + f) + \sin (\lambda ' - \Omega) \sin (\omega + f) \cos i,
\end{equation}

\noindent where $f$ is the satellite's true anomaly, and its inclination $i$ is measured from planet's orbit plane. Substituting Eq. (\ref{psi}) into (\ref{quadrupole}),
\begin{eqnarray*}
R = {\mu '}  \Bigl( {r^2 \over {r '}^3} \Bigr) \ {1 \over 2} [3 \cos^2(\lambda ' - \Omega) \cos^2 (\omega + f)+ 3 \sin^2 (\lambda ' - \Omega) \sin^2 (\omega + f) \cos^2 i\\ + 6 \cos(\lambda ' - \Omega) \cos (\omega + f) \sin (\lambda ' - \Omega) \sin (\omega + f) \cos i - 1].
\end{eqnarray*}

For practical reasons, we divide this expression into distinct terms:
\begin{equation}
\label{division}
R = { \mu ' \over {r '}^3} \ {1 \over 2} (3 R_1 + 3 R_2 \cos^2 i +6 R_{12} \cos i-R_0), 
\end{equation}
\noindent where
\begin{eqnarray*}
&R_1=r^2 \cos^2 (\lambda ' - \Omega) [ \cos^2 \omega \cos^2 f - {1 \over 2} \sin(2\omega) \sin(2f) + \sin^2 \omega \sin^2 f ],\\ 
&R_2=r^2 \sin^2 (\lambda ' - \Omega) [ \cos^2 \omega \sin^2 f + {1 \over 2} \sin(2\omega) \sin(2f) + \sin^2 \omega \cos^2 f ],\\ 
&R_{12}=r^2 \ {1 \over 4} \sin (2 \lambda ' - 2 \Omega) [\cos(2 \omega) \sin(2f) + \sin (2 \omega) \cos(2f) ],
\end{eqnarray*}
and
$$R_0=r^2.$$ 

\noindent Here we have separated the dependence on $\omega$ and $f$, in order to prepare the expression for averaging over $f$.

We average these expressions over the satellite's orbital motion by integration: $<R_i>={1 \over 2 \pi} \int_0^{2 \pi} R_i \ n_s dt$. This is most easily done by expressing $r$, $f$ and $n_s dt$ in terms of the eccentric anomaly $E$. Thus
\begin{eqnarray*}
&<r^2> = a^2 ( 1 + {3 \over 2} e^2)\\
&<r^2 \cos^2 f> = a^2 ( {1 \over 2} + 2 e^2)\\
&<r^2 \sin^2 f> = a^2 (1-e^2) / 2,
\end{eqnarray*}
and
$$<r^2 \sin(2f)> = 0.$$
Putting the averaged values back into the expressions for $R_i$:
\begin{eqnarray*}
&<R_1>=a^2 \cos^2 (\lambda ' - \Omega) [ {1 \over 2} (1 - e^2) + {5 \over 2} \ e^2 \cos^2 \omega ],\\ 
&<R_2>=a^2 \sin^2 (\lambda ' - \Omega) [ {1 \over 2} (1 - e^2) + {5 \over 2} \ e^2 \sin^2 \omega ],\\ 
&<R_{12}>=a^2 \ {1 \over 2} \sin (2 \lambda ' - 2 \Omega) \  {5 \over 4} \ e^2 \sin (2 \omega), 
\end{eqnarray*}
and
$$<R_0>=a^2 ( 1 + {3 \over 2} e^2).$$
Since both the Kozai theory (Eqs. \ref{kozaii} to \ref{kozaio}) and the lunar evection \citep{bro61} depend on double angles, we express $R$ in terms of double angles only. With these average values, Eq. (\ref{division}) can be rewritten as
\begin{eqnarray*}
&R' = R \ (2 \ r'^3 / \mu' a^2)\\ 
&= {1 \over 2} - 3 e^2 - {3 \over 2} (1-e^2) \sin^2 i \sin^2(\lambda ' - \Omega)+ {15 \over 8} e^2 \{ \ [ 1+\cos(2 \lambda ' -2 \Omega) ] \ [ 1 + \cos (2\omega) ]\\ 
&+ \cos^2 i [1 - \cos(2\lambda ' - 2\Omega)]\ [1- \cos (2 \omega) ] + 2 \cos i \sin (2 \omega) \sin (2 \lambda ' - 2 \Omega) \}\\
\\
&= {1 \over 2} - 3 e^2 - {3 \over 4} (1-e^2) \sin^2 i [1 -\cos(2 \lambda ' - 2 \Omega)]\\
&+ {15 \over 8} e^2 \{  (1+\cos^2i) [ 1 + \cos (2\omega) \cos (2 \lambda' - 2\Omega)]\\ 
&+ (1-\cos^2 i) [\cos(2\lambda ' - 2\Omega)+ \cos (2 \omega)] + 2 \cos i \sin (2 \omega) \sin (2 \lambda ' - 2 \Omega) \}.
\end{eqnarray*}
The resulting disturbing function can be divided into three pieces:
\begin{equation}
\label{threeparts}
R = R' \ { \mu ' a^2  \over 2 {r '}^3} = R_K \ ( 2 \omega) + R_I \ (2 \lambda '- 2 \Omega)  + R_E \ (2 \omega, \ 2 \lambda' - 2 \Omega)
\end{equation}
with the parts being
\begin{equation}
\label{kozair}
R_K = { \mu ' a^2  \over 2 {r '}^3} \{ {1 \over 2} - 3 e^2 - {3 \over 4} (1-e^2) \sin^2 i+ {15 \over 8} e^2 [1 + \cos^2 i + \sin^2 i \cos (2 \omega) ] \},
\end{equation}
\begin{equation}
\label{evictionr}
R_I= { \mu ' a^2  \over 2 {r '}^3} [ {3 \over 4} (1-e^2) + {15 \over 8 } e^2 ] \sin^2 i \cos( 2 \lambda' - 2 \Omega),
\end{equation}
and
$$R_E={15 \over 16} { \mu ' a^2  \over {r '}^3} e^2 \{ (1+ \cos^2 i) \cos (2 \omega) \cos (2 \lambda ' - 2 \Omega)$$ 
\begin{equation}
\label{evectionr}
+ 2 \cos i \sin (2 \omega) \sin (2 \lambda ' - 2 \Omega) \}.
\end{equation}
The first term, $R_K$, is the only one that is retained in the Kozai theory. Since it is independent of $\lambda'$, the averaging over the ``orbital motion'' of the Sun (which is, of course, only the reflection of the planet's motion) is trivial. The task reduces to:
\begin{equation}
\label{lambda}
{1 \over 2 \pi} \int_0^{2 \pi} {n_p dt \over r'^3} = {1 \over a'^3 (1-e'^2)^{3/2}}.
\end{equation}
When we put this result back into Eq. \ref{kozair}, introduce $b'=a' \sqrt{1-e'^2}$ (semiminor axis of the planet's orbit) and group the terms differently, we get:
\begin{equation}
\label{innanen}
R_K={\mu ' a^2 \over 8 b'^3} [ 2 + 3 e^2 - (3 +12 e^2 -15 e^2 \cos^2 \omega) \sin^2 i ].
\end{equation}
This is the exact form of the disturbing function obtained by \citet{inn97}, which, through Lagrange's equations, yields Eqs. (\ref{kozaii}--\ref{kozaio}).  

Now we consider $R_E$, which, if $i=0$ (i.e., $\cos i =1$), reduces to
$$R_E={15 \over 8} { \mu ' a^2  \over {r '}^3} e^2 \cos (2 \lambda ' - 2 \Omega - 2 \omega)$$
\begin{equation}
\label{brouwer}
 = {15 \over 8} { \mu ' a^2  \over {r '}^3} e^2 \cos (2 \lambda ' - 2 \varpi);
\end{equation}
\noindent this is equivalent to the evection term given by \citet{bro61} in their chapter on lunar theory. However, since we are interested in irregular satellite dynamics, we cannot restrict $i$ to be $0^\circ$ and must retain the dependence on $i$. Also, we cannot use $\varpi$ at this point, since it is not defined similarly for all satellites: $\varpi=\Omega + \omega$ for prograde bodies, but $\varpi=\Omega - \omega$ for retrograde ones. Nevertheless, we can calculate the secular trend resulting from $R_E$ in much the same way that \citet{bro61} did. We recognize that this term produces a periodic perturbation in $e$. Using the Lagrange equation for $\dot e$ \citep{dan92}, we find
$$\Bigl({de \over dt}\Bigr)_E={15 \over 8}{ \mu ' n_s a^3  \over \mu {r '}^3}  e \sqrt{1-e^2} [ (1+ \cos^2 i) \sin (2 \omega) \cos (2 \lambda ' - 2 \Omega)$$ 
\begin{equation}
\label{dedt}
- 2 \cos i \cos (2 \omega) \sin (2 \lambda ' - 2 \Omega) ].
\end{equation}

If we consider that $\lambda' \simeq n_p t$, and that it changes much faster than the satellite's orbital elements, we can integrate Eq. \ref{dedt}, getting
$$\delta e_E={15 \over 16} { \mu ' n_s a^3  \over \mu n_p {r '}^3}  e \sqrt{1-e^2} [ (1+ \cos^2 i) \sin (2 \omega) \sin (2 \lambda ' - 2 \Omega)$$ 
\begin{equation}
\label{de}
+ 2 \cos i \cos (2 \omega) \cos (2 \lambda ' - 2 \Omega) ].
\end{equation}
Since $a^3/\mu=n_s^{-2}$ and $\mu'/r'^3 \simeq n_p^2$, the factor $[({\mu ' / \mu}) ({n_s / n_p}) ({a / r'})^3]$ in Eq.~\ref{de} approximately equals $n_p/n_s=m$, so $\delta e_E$ is of order $m e$. Ignoring higher orders of $m$ for now, we can write $e^2=\bar{e}^2+2\bar{e} \delta e_E$, where the new $e$ is independent of $\lambda'$ to order $m$. Now Eq. \ref{evectionr} becomes:
$$R_E={15 \over 16} { \mu ' a^2  \over {r '}^3} e^2 \Bigl\{ 1+ {15 \over 8} {\mu ' n_s a^3  \over \mu n_p {r '}^3} \sqrt{1-e^2}$$ 
$$\times [ (1+ \cos^2 i) \sin (2 \omega) \sin (2 \lambda ' - 2 \Omega)+ 2 \cos i \cos (2 \omega) \cos (2 \lambda ' - 2 \Omega) ] \Bigr\}$$
\begin{equation}
\label{evectione}
\times \Bigl\{ (1+ \cos^2 i) \cos (2 \omega) \cos (2 \lambda ' - 2 \Omega)+ 2 \cos i \sin (2 \omega) \sin (2 \lambda ' - 2 \Omega) \Bigr\};
\end{equation}
\noindent here the first $\{ \}$ arises from our changes in $e$. We seek the secular effects arising from evection, so we are interested solely in those parts of (\ref{evectione}) that survive integration over $\lambda'$ from 0 to $2  \pi$; these are those terms that contain $\cos^2 (2 \omega) \cos^2 (2 \lambda ' - 2 \Omega)$ or $\sin^2 (2 \omega) \sin^2 (2 \lambda ' - 2 \Omega)$. Such terms, when averaged over $\lambda'$ (cf. Eq. \ref{lambda}), produce factors of ${1 \over 2} \cos^2 (2 \omega)$ or ${1 \over 2} \sin^2 (2 \omega)$, respectively. Since we are interested only in the leading terms coming from evection, we can arguably ignore the ellipticity of the planet's orbit. Accordingly, the averaged $R_E$ is
$$< R_E >={225 \over 128} \Bigl( {\mu'\over a'^3} \Bigr)^2 \Bigl( {a^3 \over \mu} \Bigr) {n_s \over n_p} a^2 e^2 \sqrt{1-e^2} (1+\cos^2 i) \cos i [\cos^2 (2 \omega) + \sin^2 ( 2 \omega)].$$ 
The dependence on $\omega$ obviously vanishes when the last two terms are added, and the lead factor $(a / a'^2)^3$ can be given instead by the mean motions. Hence
\begin{equation}
\label{average}
<R_E>={225 \over 128} m \ n_p^2 a^2 e^2 \sqrt{1-e^2} (1+\cos^2 i) \cos i.
\end{equation}
The extra $\dot \omega$ (Danby 1992) arising from this term is 
$$\Bigl( {d\omega \over dt} \Bigr)_E={225 \over 128} m n_p^2 {n_s a^3 \over \mu} \cos i \Bigl\{ [ 2 (1-e^2) - e^2](1+\cos^2 i) + e^2 (1+3 \cos^2 i) \Bigr\}.$$ 
Replacing ${n_p n_s a^3 / \mu}$ with $m$ and grouping the terms, we get
\begin{equation}
\label{dwdt}
\Bigl( {d\omega \over dt} \Bigr)_E={225 \over 64} m^2 n_p \cos i ( 2 -\sin^2 i -  e^2 ).   
\end{equation}
\noindent Similarly,
\begin{equation}
\label{dodt}
\Bigl( {d\Omega \over dt} \Bigr)_E=-{225 \over 128} m^2 n_p e^2 (4-3 \sin^2 i).
\end{equation}
We note that ${\dot \Omega}_E$ has $e^2$ as the leading term in eccentricity, whereas ${\dot \omega}_E$ is non-zero even for circular orbits. In the case when $i=0$ and $e=0$, ${\dot \varpi}_E={\dot \omega}_E + {\dot \Omega}_E = (225/32) \ m^2 n_p$, which is the second term in Eq. \ref{moon}, and comes from classical lunar theory. Therefore, it is clear that evection is by far the largest correction to any purely secular theory, such as Kozai's. Although Eq. \ref{dwdt} describes the only evection-related secular perturbation that affects circular orbits, it does not represent the full effect of the evection term in the disturbing function (Eq. \ref{evectionr}). To derive Eqs. \ref{dwdt} and \ref{dodt} we used only the perturbation that a satellite's eccentricity suffers from evection. The fact that Eq. \ref{evectionr} depends on $\Omega$ indicates that we should take into account that the inclination, too, suffers a periodic perturbation. 

The equivalent of Eq. \ref{dedt} for inclination is:
$$\Bigl({di \over dt}\Bigr)_E=- {15 \over 8} { \mu ' n_s a^3  \over \mu {r '}^3} {e^2 \over \sqrt{1-e^2}}{1 \over \sin i} \times $$ 
$$\times \{ (1+\cos^2 i)[ \cos(2 \omega) \sin (2 \lambda ' - 2 \Omega)+ \cos i \sin (2 \omega) \cos (2 \lambda' - 2 \Omega)]$$
\begin{equation}
\label{didt}
- 2 \cos i [\sin (2 \omega) \cos (2 \lambda' - 2 \Omega)+\cos i \cos (2 \omega) \sin (2 \lambda ' - 2 \Omega)]\}.
\end{equation}
Integrating as we did for Eq. \ref{dedt}, 
$$\delta i_E=- {15 \over 16} {n_s \over n_p}{ \mu ' a^3  \over \mu {r '}^3} {e^2 \over \sqrt{1-e^2}}{1 \over \sin i} \times $$ 
$$\times \{ (1+\cos^2 i)[ - \cos(2 \omega) \cos (2 \lambda ' - 2 \Omega)+ \cos i \sin (2 \omega) \sin (2 \lambda' - 2 \Omega)].$$
\begin{equation}
\label{di}
- 2 \cos i [\sin (2 \omega) \sin (2 \lambda' - 2 \Omega)-\cos i \cos (2 \omega) \cos (2 \lambda ' - 2 \Omega)]\}.
\end{equation}
Since the inclination enters Eq. \ref{evectionr} only through $\cos i$, it is useful to simplify the last expression and convert it to $\delta (\cos i)_E$:
$$\delta (\cos i)_E=- {n_s \over n_p}{ \mu ' a^3  \over \mu {r '}^3} {15 \over 16} {e^2 \over \sqrt{1-e^2}} \sin^2 i \times$$ 
\begin{equation}
\label{dcosi}
\times \{ \cos(2 \omega) \cos (2 \lambda ' - 2 \Omega) + \cos i \sin (2 \omega) \sin (2 \lambda' - 2 \Omega)\}.
\end{equation}
Putting (\ref{dcosi}) back into (\ref{evectionr}) and integrating the resulting expression in the same manner as the term in Eq. \ref{evectione}, we get
\begin{equation}
\label{averagei}
<R_{E'}>=- {225 \over 256} m \ n_p^2 a^2 {e^4 \over \sqrt{1-e^2}} \cos i \sin^2 i.
\end{equation}
The resulting $\dot \omega$ and $\dot \Omega$ caused by this term are:
$$\Bigl( {d \omega \over dt} \Bigr)_{E'}=-{225 \over 256} m n_p^2 {n_s a^3 \over \mu} \cos i\{(4 e^2 + {e^4 \over 1 - e^2}) \sin^2 i - {e^4 \over 1-e^2} (2 \cos^2 i - \sin^2 i)\}$$
\begin{equation}
\label{dwdti}
=-{225 \over 128} m^2 n_p \cos i (2 e^2 \sin^2 i - {e^4 \over 1-e^2} \cos (2 i)),
\end{equation}
and
\begin{equation}
\label{dodti}
\Bigl( {d \Omega \over dt} \Bigr)_{E'}=-{225 \over 256} m^2 n_p {e^4 \over 1-e^2} (2-3 \sin^2 i).
\end{equation}
Several characteristics of these terms are important. Both (\ref{dwdti}) and (\ref{dodti}) are zero for circular, planar orbits, so they make no contribution to (\ref{moon}). Likewise, $\dot {\varpi}_{E'}$ is zero for all orbits with $i=0$, regardless of eccentricity. Since the leading term in Eq. \ref{dwdti} is $e^2 \cos i \sin^2 i$, this expression (for a satellite with $e=0.5$ and $i=45^{\circ}$) is an order of magnitude smaller than the leading evection term (Eq. \ref{dwdt}).

The only term in Eq. \ref{threeparts} that we have not addressed yet is $R_I$, or the ``nodal evection'' term. Since it does not depend on $\omega$, it has no effect on the eccentricity, but only on the inclination (see the Lagrange equations in Danby 1992). The resulting perturbation is:
\begin{equation}
\label{didti}
\Bigl({di \over dt}\Bigr)_I= - { \mu ' n_s a^3  \over \mu {r '}^3} {1 \over \sqrt{1-e^2}} [ {3 \over 4} (1-e^2) + {15 \over 8 } e^2 ] \sin i \sin( 2 \lambda' - 2 \Omega).
\end{equation}
When integrated, this perturbation produces the following periodic term in $\sin^2 i$:
\begin{equation}
\label{dsinii}
\delta (\sin^2 i)_I= {n_s \over n_p}{ \mu ' a^3  \over \mu {r '}^3} {1 \over \sqrt{1-e^2}} [ {3 \over 4}(1-e^2) + {15 \over 8} e^2 ] \cos i \sin^2 i \cos( 2 \lambda' - 2 \Omega).
\end{equation}
Once (\ref{dsinii}) is substituted into (\ref{evictionr}) and the resulting expression is integrated over $\lambda' $, we get
\begin{equation}
\label{average_evi}
<R_I>= {9 \over 64} \ m n_p^2 a^2 {1 \over \sqrt{1-e^2}} (1 + 3 e^2 + {9 \over 4} e^4) \cos i \sin^2 i . 
\end{equation}
Using Lagrange's equations, we can calculate the secular effect of nodal evection: 
$$\Bigl( {d \omega \over dt} \Bigr)_I= {9 \over 64} \ m^2 n_p \Bigl[ \Bigl(6 + 9 e^2 + {1 + 3 e^2 + (9/4) e^4 \over 1-e^2}\Bigr) \cos i \sin^2 i$$
$$ - \ {1 + 3 e^2 + (9/4) e^4 \over 1-e^2} \cos i (2 \cos^2 i - \sin^2 i) \Bigr]$$  
\begin{equation}
\label{dwdt_evi}
= {9 \over 32} \ m^2 n_p \cos i \Bigl[ (3 + {9 \over 2} e^2 ) \sin^2 i -{(1 + 3 e^2/2)^2 \over 1-e^2} \cos (2 i) \Bigr], 
\end{equation}
\noindent and
\begin{equation}
\label{dodt_evi}
\Bigl( {d \Omega \over dt} \Bigr)_I= {9 \over 64} \ m^2 n_p {(1 + 3 e^2/2)^2 \over 1-e^2} (2 - 3\sin^2 i). 
\end{equation}
While the numerical coefficients in front of the nodal evection term are much smaller than that for apsidal evection, nodal evection still needs to be addressed if we want to determine the rate of nodal regression accurately. Eqs. \ref{dwdt_evi} and \ref{dodt_evi} indicate that $\dot \varpi$ is not affected by nodal evection if the orbit has no inclination, while there is always some contribution to $\dot \Omega$.  When $e=0$ and $i=0$, $\dot \Omega_I=9/32 \ m^2 n_p$, which is identical to the second term  in the expansion in $m$ of the lunar nodal regression rate (the first being $-3/4 \ m n_p$; Brouwer \& Clemence 1961, see also Saha \& Tremaine 1993).

Until now, we have determined the exact form of all terms of order $m^2 n_p$ or lower that enter the expression for the precessions of $\omega$ and $\Omega$. We have, however, neglected any terms containing higher powers of $m$, which are bound to arise in multiple places. Eq. \ref{moon} clearly demonstrates that $\dot \varpi$ converges very slowly over powers of $m$, implying that many terms have to be included to achieve an acceptable accuracy for satellites with $m$'s comparable to the Moon's (0.075). To illustrate this, Fig. 1 shows how the apsidal precession rate depends on the satellite's eccentricity in various theories and in direct numerical integration. The planet's eccentricity is taken to be very low (0.01) to avoid the effects arising from the octupole term (Section 4). Fig. 1 indicates that, while evection comprises about 2/3 of the discrepancy between the Kozai theory and the numerical integration, the accuracy of an analytical theory using only terms up to $m^2$ is limited, especially for low-eccentricity orbits.

Even though the direct derivation of terms with higher powers of $m$ appears to be very difficult if we are to use the same method as in this section, we can make some reasonable assumptions about their form. First, we note that all the $m^2$ terms in $\dot \omega$ contained $\cos i$, making them symmetric rather than antisymmetric with respect to the direction of orbital motion. Comparing our result with Eq. \ref{moon}, we note that the variable $m$ actually has a different meaning in our approach from that in lunar theory. In Eq. \ref{moon}, $m$ can be both positive and negative, to accomodate prograde and retrograde orbits, respectively. In the Kozai theory, just like in the present paper, the inclination is dealt with explicitly, and $m$ (as well as the mean motion of the satellite) is always considered to be positive. So, in Eq. \ref{moon} the leading term changes sign for retrograde satellites, while the $m^2$ term is always positive. Eq. \ref{kozaiw} states that $\dot \omega$ is always positive (at least for low-$i$ orbits); however, owing to the different definition of $\varpi$ for retrograde orbits ($\Omega -\omega$), the motion of the line of apsides is negative. Likewise, the $m^2$ term in Eq. \ref{moon} is always positive, which requires the presence of a factor $\cos i$ in the $\dot \omega$ term causing it (Eq. \ref{dwdt}). We therefore propose that every subsequent term in $\dot \omega$ that contains $m$ to a power $n$, should contain $\cos i$ to the power of $n-1$, in order to make the terms alternately symmetric and antisymmetric. 

Figure 1 indicates that terms in $\dot \varpi$ beyond the evection itself decrease with increasing $e$, so we need to address this dependence. We observe that the secular-plus-evection curve intersects both the Kozai curve and the numerical values at approximately $e=0.8$. Based on this, we will postulate that higher-order terms in $m$ behave like $(1-  3 e^2/2)$. Note that $\dot{\varpi}$ coming from evection has the same dependence on $e$ in the planar case (Eqs. \ref{dwdt} and \ref{dodt}). Unlike the dependence of higher-order terms on inclination discussed in the paragraph above, this approximation is more pragmatic and does not pretend to be exact.

So our final equations for $\dot \omega$ and $\dot \Omega$ are
\begin{equation}
\label{finaldwdt}
\Bigl( {d\omega \over dt} \Bigr)=\Bigl( {d\omega \over dt} \Bigr)_K+\Bigl( {d\omega \over dt} \Bigr)_E+\Bigl( {d\omega \over dt} \Bigr)_{E'}+\Bigl( {d\omega \over dt} \Bigr)_I+\Bigl( {d\omega \over dt} \Bigr)_L
\end{equation}
and
\begin{equation}
\label{finaldodt}
\Bigl( {d\Omega \over dt} \Bigr)=\Bigl( {d\Omega \over dt} \Bigr)_K+\Bigl( {d\Omega \over dt} \Bigr)_E+\Bigl( {d\Omega \over dt} \Bigr)_{E'}+\Bigl( {d\Omega \over dt} \Bigr)_I,
\end{equation}
where the $\dot \omega$ terms with subscripts $K$, $E$, $E'$ and $I$ are given by Eqs. \ref{kozaiw}, \ref{dwdt}, \ref{dwdti} and \ref{dwdt_evi}, respectively. Likewise, Eqs. \ref{kozaio}, \ref{dodt}, \ref{dodti} and \ref{dodt_evi} define the $\dot \Omega$ terms with respective subscripts $K$, $E$, $E'$ and $I$. We define $\dot \omega_L$ as:
\begin{equation}
\label{dwdtl}
\Bigl( {d\omega \over dt} \Bigr)_L= n_p (1- {3/2} \ e^2) \sum_{j=3}^{10} C_j \ m^j \cos^{j-1}i, 
\end{equation}
where the $C_j$ are taken from \citet{tis94}, and are very similar (but not identical) to the coefficients in Eq. A1 of \citet{saha93}. To obtain a closed system of equations that can be evolved numerically, we also need to include expressions for the secular evolution of $i$ and $e$ from the Kozai theory (Eqs. \ref{kozaii} and \ref{kozaie}). Evection induces only short-period variations in eccentricity which were fully accounted for during the averaging process, and has no consequences for the long-term evolution of the eccentricity (except for distant prograde orbits that are unstable; Nesvorn\' y et al. 2003). The $\dot \varpi$ predicted by Eqs. (\ref{finaldwdt}) and (\ref{finaldodt}) for a planar, prograde case is plotted in Fig. 1, and its agreement with numerical integrations is quite satisfactory. In the next section, we compare the predictions of the present theory to the direct numerical integrations for a wide range of mean motions, eccentricities and inclinations.    

\section{Comparison with Numerical Integrations}

Our first step in evaluating how closely our analytical model represents the behavior of real satellites is to compare it to a set of directly integrated test particles. In order to classify possible deviations of the model from the observed dynamics, we will first look at orbits that either have high eccentricity or high inclination, but never both. Only after that will we apply the model to the real satellites, which often exhibit both. Recall that in the conventional Kozai theory, $(1-e^2)^{1/2}\cos i$ (the orbit's angular momentum that lies normal to the reference plane) is conserved (see Holman et al. 1997, Carruba et al. 2002). Thus changes in $e$ and $i$ are coupled. Hence, throughout this chapter, we will characterize bodies by their ``minimum inclination'' and ``maximum eccentricity''. By this we are referring to the values reached at one extreme of the Kozai cycle, when $\omega=90^{\circ}$. This way the orbits are uniquely defined in a secular model which ignores octupole and higher terms, since it depends only on $a$, $e$, $i$ and $\omega$ (the inclusion of the octupole term brings in a dependence on $\varpi$; Yokoyama et al. 2003). We prefer these ``extreme elements'' over the mean ones since two irregular satellites with quite different orbits can have the same mean elements. Also, these extreme elements are used as initial conditions in all the integrations shown in this chapter. If the Kozai resonance is possible for a particle's combination of $a$, $e$ and $i$, an initial $\omega=90^{\circ}$ will put it in the libration region. This way, all of our test particles above the critical inclination will have librating $\omega$, enabling us to find the exact location of the boundary.

The first group of integrations deals with the secular behavior of test particles having very low minimum inclination ($5^{\circ}$, or $175^{\circ}$ for retrograde cases) and maximum eccentricities varying from 0 to almost 1. In all secular theories, Eqs. \ref{kozaii} and \ref{kozaie} were used for the evolution of inclination and eccentricity, respectively. All of the continuous lines in the figures were obtained through advancing the relevant secular equations by a Burlisch-Stoer-type numerical integrator over multiple precession periods, and then computing the average precession rate. The discrete points show the results of a direct numerical simulation for the same parameters. Just as in Fig. 1, the eccentricity of the planet (``pseudo-Jupiter'') in the symplectic intergration was taken to be only 0.01 to avoid any interference from the octupole term (shown in Sec. 4 to be proportional to $e_p$).

It is clear from Fig. 2a that our model is a very good approximation to the secular behavior of a prograde, low-$i$ satellite for all values of eccentricity. While the addition of higher-order terms (Eq. \ref{dwdtl}) was needed to accurately describe the motion of the line of apsides, evection alone is sufficient for a satisfactory description of nodal regresion for all $e$. Fig. 2b shows that our model of apsidal precession gives excellent results for retrograde satellites, too. However, the precession rate of the node differs from the one predicted for particles with $e>0.5$ (although our model is still more accurate than Kozai's as long as $e<0.9$). The most likely cause of this disagreement is the existence of higher-order terms in $m$ that affect the nodal precession. This hypothesis is supported by Fig. 2c, in which the theory and numerical simulation of the nodal precession noticeably diverge already at $e>0.4$. In contrast, our predictions for the apsidal precession are still reasonably accurate, and the slight shift between two curves is most likely a consequence of $C_j$'s with $j~>~10$, that are not included in expression (\ref{dwdtl}). Unfortunately, since these ``phantom terms'' in $\dot \Omega$ vanish if either $e=0$ or $i=0$ (but not when $i=180^{\circ}$), there is no easy way to synthesize them on the basis of lunar theory, as we did with apsidal terms in Eq. \ref{dwdtl}. This inability of our empirical approximation to describe accurately the nodal motion of distant and eccentric retrograde orbits is irrelevant for the discussion of secular resonances involving only the longitude of pericenter (Sections 4 and 5). On the other hand, as we shall see, the calculation of the boundary of Kozai resonance is affected, with the model giving us erroneous results for the secular behavior of at least one known irregular satellite (see our later discussion of Fig. 4b).

Figures 3a-c test the behavior of our model with changing minimum inclination. The maximum eccentricity in all three panels is 0.2. In all three panels we see that at a certain inclination the average precession rates of the line of apsides and of the line of nodes converge and stay equal to each other at all higher inclinations. When those two rates are identical, the argument of pericenter librates around $90^\circ$ or $270^\circ$; this behavior is usually known as the Kozai resonance (see Sec. 1). It is interesting that the Kozai resonance is not achieved by both secular solutions at the same inclination. In Fig. 3a the Kozai resonance is reached by our model at a minimum inclination of about $47^\circ$, while the Kozai model, as expected, predicts the boundary to be at $39.2^\circ$ (in Kozai theory this result is independent of the mean motion). The numerical simulation, however, agrees closely with our model (the observed slight divergence within the libration region arises because the points actually show the average of inclinations at which $\omega$ passes through the libration center, regardless of the direction). This higher threshold for the Kozai resonance among distant prograde objects led to erroneous early predictions based on purely secular models that Siarnaq (formerly S/2000~S3) should be in Kozai resonance \citep{vas01}; direct numerical integrations showed it to be in the secular resonance instead \citep{cuk02a, nes03}. In Figs. 3b and 3c, the Kozai resonance is reached at inclinations larger (i.e., closer to $180^\circ$) than $140.8^\circ$ by both the numerical simulation and our model. Although our model puts this threshold at slightly larger inclination than the simulation, it still approximates the numerical results much better than does the traditional Kozai model. 

Figures 2 and 3 suggest that our model fairly accurately predicts the precessional rate of the line of apsides for all the cases tested, and that it can predict the motion of the line of nodes for all prograde orbits and for those retrograde orbits that have low eccentricity. Figs. 3a-c show that the boundary between $\omega$'s libration and circulation is not fixed but varies with different mean motions. The shift in this boundary reduces the range of inclinations over which prograde orbits can be in Kozai libration, while it expands the range of retrograde inclinations that allow libration. 

To identify the region in which Kozai resonance is possible, we determine the inclination at which circular orbits switch from circulation to libration as a function of the orbital period (measured by $m$). To make the plot continuous, we define inclination to be always smaller than $90^\circ$, with $m$ taken to be positive for prograde bodies and negative for retrograde ones. The boundary is calculated by substituting $e=0$ and $\omega=90^\circ$ in the separate terms in Eq. \ref{finaldwdt}, which is then equated to zero:
$${d\omega \over d\tau}=2-5 \sin^2 i + {225 \over 64} m \cos i(2-\sin^2 i)$$
\begin{equation}
\label{boundary}
 + {9 \over 32} m \cos i (3 \sin^2 i - \cos(2 i))  + \sum_{j=3}^{10} C_j \ m^{j-1} \cos^{j-1}i=0 \ . 
\end{equation}
The classical result that Kozai resonance occurs for $i > 39.2^\circ$ (or $i < 140.8^\circ$) comes from the first two terms on the right ($\sin^2 i=2/5$). Equation \ref{boundary} is solved iteratively for $i$, by first assuming that $\sin^2 i=0.4$, and then finding successive approximations to $i$ using
$$i_{n+1}=\arcsin \{ {2 \over 5}+{45 \over 64} m \cos i_n(2-\sin^2 i_n)$$
\begin{equation}
\label{iteration}
 + {9 \over 160} m \cos i_n [3 \sin^2 i_n - \cos(2 i_n)]  + {1 \over 5}\sum_{j=3}^{10} C_j \ m^{j-1} \cos^{j-1}i_n \}^{1/2} \ . 
\end{equation}
 This expression converges well for $m$ values that are typical of the known irregular satellites (-0.2 $\rightarrow$ 0.1); since $m$ and $\cos i$ always appear together in Eq. \ref{iteration}, the expression does not need to be modified now that we switch to a new convention in which retrograde orbits have $m<0$ and $i < 90^\circ$. Fig. 4a plots the location of the Kozai resonance as defined by this procedure.

 Since circular orbits, strictly speaking, should exhibit no Kozai cycle in inclination over the precession (or libration) period of $\omega$, this inclination is both a minimum and an average at the same time. So we also plot, as individual points, the $m$'s and minimum inclinations of some known irregular satellites that lie in the same region. These minimum inclinations were determined on the basis of a 300,000-year symplectic integration for each satellite. For these integrations we used our home-made orbital integration software, which follows the standard algorithm devised by \citet{wis91}. The initial conditions were obtained from JPL's Horizons ephemeris service on December 29, 2003, in the form of the osculating elements for the four giant-planet barycenters and the irregular satellites themselves for midnight, September 24, 2003. The \underline{minimum inclination} is defined as the average inclination to the planet's orbital plane for times when $\cos \omega < 0.1$ and the eccentricity is larger than the average value for the integration. Different symbols represent the satellites of each of the four giant planets (see figure caption); those satellites that are found to have librating $\omega$'s are also marked by large triangles, as well as identified. The satellites of Jupiter that were discovered in 2003 are not included in this plot, as their orbits were not well constrained when the initial conditions were generated, although they are very likely to fall into the already known orbital clusters \citep{she03}. An exception was made for S/2003 J20, which is very unlikely to belong to any of the known groups, and is therefore probably relevant to the question of the overall orbital distribution of the originally captured bodies.

Two features of Fig. 4a require comment. First, clearly our theoretical prediction for the circular case poorly describes the behavior of the real satellites. More Kozai librators are found slightly below the line than above it; also, a couple of Neptunian retrograde irregulars that are above the line are not thought to be Kozai librators. The reasons for this discrepancy are that almost all of the librating satellites have substantial eccentricities, and that the boundary between the two types of secular behavior shifts with increasing eccentricity. Just from the distribution of known librators it is possible to speculate that the boundary moves down for more eccentric prograde moons, and up for eccentric retrogrades. The only satellite for which this low-$e$ approximation is accurate is Euporie (formerly S/2001 J10), which indeed does have low eccentricity ($e_{max}=0.17$). The other important feature of this plot is that the boundary goes through a local minimum in inclination close to Euporie's position ($m=-0.13$); for more distant retrograde orbits, the circulation-libration transition happens at increasingly larger inclinations. This change in behavior apparently occurs because, at more negative $m$'s, terms of order 3 and higher in Eq.~\ref{iteration} become more important than the pair coming from evection.

To account fully for the effects of eccentricity on the position of the Kozai boundary, we need to carry out a full integration of the equations affecting $\omega$'s motion (Eqs. \ref{kozaii}, \ref{kozaie} and \ref{finaldwdt}). Although Eq. {\ref{finaldodt} is decoupled from the other three ($\Omega$ cannot influence other elements in the quadrupole-limited theory), we evolve it alongside the other three for comparison.  We do this for three sets of test particles that form a dense grid in $m$ and minimum inclination (the cell size is 0.01 $\times 1^\circ$), with the sets having maximum eccentricities of 0.2, 0.4 and 0.6. The planet's parameters are those chosen for Figs. 2 and 3. After the particles' orbits are advanced over tens of precession periods with a Burlisch-Stoer-type integrator, the average precession periods of $\omega$ and $\Omega$ are computed for each particle. Then, for each $m$, we seek the particle with the lowest inclination for which $\dot{\omega}  < 0.1 \dot{\Omega}$ (a relative criterion is required since close-in moons always have longer secular periods than the more distant ones), and we use this to indicate the onset of the Kozai resonance. These lowest minimum-inclinations for which $\omega$ is librating are connected by continuous lines in Fig. 4b. 

As expected, the boundary for the Kozai resonance among the prograde satellites moves to lower inclinations with increasing eccentricity. Now all of the known prograde librators are either above, or less than a degree below, the boundary for their eccentricity range (the way that boundaries were computed can lead to critical inclinations being overestimated by as much as one degree). No circulating satellites are found to have minimum inclinations above the boundary. We conclude that our model predicts the position of the separatrix between circulation and libration among prograde moons accurately enough for the purpose of discussing their overall distribution and behavior. It is also noteworthy that all but one (S/2003 U3, the orbit of which still has to be confirmed) among the prograde librators are fairly close to the circulation-libration boundary, which may offer some clues about their origin and past evolution (Section 6). 

On the other hand, the agreement between the Kozai boundary predicted by our theory and the behavior of high-$i$ retrograde satellites is poor, especially for the more distant bodies. Instead of shifting to higher $i$ as $e$ increases (as might be suggested by the positions of the real satellites on the plot), the computed boundary appears to do exactly the opposite. The cause of this discrepancy is clearly illustrated in Fig. 2c. Our model fails to predict the nodal precession rate of the distant and eccentric retrogrades with any accuracy, while it does much better when applied to the motion of the line of apsides. Since the Kozai resonance deals with the behavior of the argument of pericenter (defined as the angle between the ascending node and the periapse), our model should be expected to produce major errors when trying to predict its location. Fig. 2c indicates that, according to our  model, the precession of the line of nodes (of a distant, retrograde satellite) slows down with increasing eccentricity, while the numerical simulation suggests that it actually accelerates. Since the Kozai resonance happens because of a reduced apsidal precession rate with increasing inclination (Fig. 3c) and, ultimately, its synchronization with the nodal precession rate, the inclination at which this synchronization arises must depend on the speed at which the nodes move. Fig. 3c suggests that a faster nodal precession would lead to the Kozai resonance starting at a higher inclination (by ``higher'' here we mean ``closer to $90^\circ$''), and vice versa. Therefore, the lines plotting the circulation-libration boundary for retrograde satellites with higher eccentricities should be consistently above the zero-$e$ curve shown in Fig. 4a. So the upturn seen at $m=-0.13$ in Fig. 4a should happen at higher $i$ and be even more pronounced for larger $e$. It is easy to anticipate that for a maximum eccentricity of about 0.7, the boundary should pass between the positions of the new Neptunian moons S/2002 N4 (labeled on the plot) and S/2003 N1 (the box directly below N4), which have similar orbits except that the former is thought to be a librator whereas the latter's $\omega$  circulates. Of course, the present-day orbits for these objects might differ from the true ones, but this does not make the numerical simulation which predicts that these two sets of initial conditions lie on opposite sides of the boundary be any less accurate. Having this in mind, we remark that the only two retrograde moons that are thought to be Kozai librators are also not far from the true (if not the modeled one) circulation-libration boundary, just like their more numerous prograde cousins.

We now identify the position of the secular resonance, defined as $\dot{\varpi}-\dot{\varpi}_{planet}=0$. This is shown in Fig. 5, which is generated from the same set of integrations as Fig. 4b. The shape of the secular resonance was found by locating, for each set of particles with the same $m$ and $e$, two test particles on neighboring inclinations that have their lines of apsides precessing in opposite directions (where $\varpi=\Omega+\omega$ for prograde, and $\varpi=\Omega-\omega$ for retrograde particles). The critical inclination for the secular resonance was then found as an intermediate value between the inclinations of this pair at which $\dot{\varpi}=0$ (the dependence of $\dot{\varpi}$ on $i$ was assumed to be locally linear).  Strictly speaking, the condition that $\dot{\varpi} \simeq 0$ is a necessary, but not sufficient, for the secular resonance's existence, and the apsidal locking itself cannot be described by a purely quadrupole theory. So the lines in Fig. 5 should only be considered as potential locations of the secular resonance, rather than the places where apsidal locking always happens. The feasibility and strength of this apsidal locking is discussed in the next section. Here we will merely note that the positions of the four resonant (or pseudo-resonant) objects -- Pasiphae ($e_{max}=0.50$), Sinope ($e_{max}=0.32$), Stephano ($e_{max}=0.32$) and Siarnaq ($e_{max}=0.49$) -- are in good agreement with the location of the secular resonance computed from our model. Small differences between our theory and direct integration are expected since we have ignored the octupole terms that are vital for the dynamics of resonant objects \citep{yok03}. A possible correlation between the secular resonance's position and the overall distribution of the known irregular satellite groups is discussed in Section 6.
 
\section{Dynamics of Secular Resonance}

Figure 5 locates those positions in orbital-element space where the precession rate of the longitude of pericenter is zero, as computed by a numerical integration of Eqs. \ref{finaldwdt} and \ref{finaldodt}. For some time it has been known that the apsides of a slowly precessing irregular satellite's orbit can become locked in a resonance with the planet's perihelion \citep{whi93, saha93}. Figure 6 displays the evolution of the resonant argument $\Psi=\varpi-\varpi_{planet}$ for a) Pasiphae, b) Sinope, c) Siarnaq and d) Stephano. These 300,000-yr simulations were carried out using a symplectic integrator, with the same initial conditions as in the previous section. It is obvious that the medium-term behavior of the resonant argument is different for each of the four objects. Pasiphae's resonant argument librates around $\Psi=\pi$ for the integration's entire length. Sinope's pericenter alternates between circulation and libration; the center of libration is the same as Pasiphae's, but the amplitude is much larger. Siarnaq also exhibits an elaborate but periodic mixture of circulation and large-amplitude libration, only now the center of libration is $\Psi=0$. Finally, Stephano's pericenter simply circulates, albeit with a very long period (nearly 200,000 yrs).

This diversity of behavior, including the librations, cannot be explained by a quadrupole-only theory (such as ours), since the quadrupole term has no dependence on the orientation of the planet's line of apsides (cf. Eq. \ref{innanen}). \citet{yok03} have derived the next two terms (in $a/a'$) of the disturbing function for satellites with arbitrary $e$ and $i$, and have found that the octupole term (i.e., the one that contains $(a/a')^3$) is responsible for apsidal locking. The complete octupole term \citep{yok03} is:
$$R_Y={c_Y \over 64} [ (-3+33 \cos i + 15 \cos^2i-45 \cos^3i) b_1 \cos(\varpi_{Sun}-\Omega+\omega)$$ 
$$+(-3-33 \cos i + 15 \cos^2i+45 \cos^3i) b_1 \cos(\varpi_{Sun}-\Omega-\omega)$$ 
$$+(15-15 \cos i - 15 \cos^2i+15 \cos^3i) b_2 \cos(\varpi_{Sun}-\Omega+3 \omega)$$ 
\begin{equation}
\label{octupole}
+(15+15 \cos i - 15 \cos^2i-15 \cos^3i) b_2 \cos(\varpi_{Sun}-\Omega-3 \omega)],
\end{equation}
where
\begin{equation}
\label{yokoyamac}
c_Y={\mu' a^3 \over a'^4} {e_p \over (1-e_p^2)^{5/2}}, 
\end{equation}
\begin{equation}
\label{yokoyamab1}
b_1=-(5/2)e-(15/8)e^3 
\end{equation}
and
\begin{equation}
\label{yokoyamab2}
b_2=-(35/8)e^3.
\end{equation}
Yokoyama et al. were the first to be able to derive this octupole term for arbitrary $e$ and $i$, but a simple version of this term has been featured in lunar theories. If we substitute $\cos i=1$ (i.e., $i=0$) in Eq. {\ref{octupole} and ignore powers of $e$ and $e_p$ beyond linear, all the terms in Eq. \ref{octupole} vanish except for the second, which becomes 
\begin{equation}
\label{octu_moon}
R_{Y,M}=-{15 \over 16} {\mu' a^3 \over a'^4} e e_p   \cos (\varpi_{Sun}-\Omega-\omega) = - {15 \over 16} n_p^2 a^2 {a \over a'} e e_p \cos (\varpi-\varpi_{Sun}).
\end{equation}
\citet{bro61} mention expression \ref{octu_moon} as the most important term in the lunar disturbing function that produces a perturbation independent of $m$. This is to be expected, since we have seen in Sec. 2 that $m$ arises when short-period perturbations are averaged, while the octupole term is purely secular. Apart from this interesting feature, the octupole term is a very minor perturbation to the lunar motion, because the factor $a/a'$ is very small ($2.5 \times 10^{-3}$) for the Moon's orbit.

\citet{yok03} correctly point out that the octupole terms become important only if the argument of one of their cosine functions changes very slowly. Near secular resonance, the arguments in one or the other of the first two terms actually do become very slow for retrograde and prograde objects, respectively (remember that $\varpi=\Omega+\omega$ for prograde orbits, but $\varpi=\Omega-\omega$ for retrograde ones). In order for the arguments of the third and fourth terms to become stationary, a satellite must have a very slowly precessing argument of pericenter, as well as a sluggish $\Psi$. Such a resonance may actually be possible close to the boundary of Kozai libration, but these terms are irrelevant for our discussion of the secular resonance (for a detailed analysis of the dynamics close to the circulation-libration boundary, see Carruba et al. 2004). Therefore the portion of the octupole term relevant for the secular resonance \citep{yok03} is
\begin{equation}
\label{resonant}
R_r=-{c_Y \over 64} (-3 - 33 | \cos i | + 15 \cos^2 i + 45 | \cos^3 i |) \ b_1 \cos \Psi,
\end{equation}
where it is implied that $\Psi=\Omega+\omega-\varpi_p$ for prograde, and $\Psi=\Omega-\omega-\varpi_p$ for retrograde objects (where $\varpi_p=\varpi_{Sun}+\pi$ is the planet's longitude of pericenter). The eccentricity and inclination of the satellite that appear in Eq. \ref{resonant} represent average values over many secular (i.e., Kozai) periods, since $\dot {\varpi} << \dot {\omega}$, by definition, for objects in the secular resonance.

We will now describe how Eq. \ref{resonant} may explain some of the diverse behavior among the known resonant and near-resonant satellites. First, we point out that the relevant parameter for resonant-locking is not the absolute magnitude of the octupole term, but how it compares to other perturbations, most notably the quadrupole term. Eqs. \ref{kozair}--\ref{evictionr} show that all parts of the quadrupole term contain the tidal factor $\mu' a^2 /a'^3$, which effectively determines its strength. So, to assess the resonant term's relative importance, we divide Eq. \ref{resonant} by $\mu' a^2/ a'^3$. Assuming $e_p << 1$, the resulting expression is
\begin{equation}
\label{rr}
R'_r=-e_p {a \over a'} S (\cos i) b_1 (e) \cos \Psi,
\end{equation}
where $S (\cos i)$ is 
\begin{equation}
\label{S}
S(\cos i)=(-3 - 33 | \cos i | + 15 \cos^2 i + 45 | \cos^3 i |)/64. 
\end{equation}
We see that the resonant term is proportional to the planet's eccentricity, meaning that it will vary in strength as $e_p$ changes. Fig. 6b suggests that Sinope's episodes of libration and circulation alternate with a period of roughly 50,000 yrs. We recall that secular changes in Jupiter's eccentricity have a similar period \citep{md99}. We therefore suggest that resonant-locking is drivern by $e_p$. To illustrate this correlation between the behavior of Sinope's resonant argument and Jupiter's eccentricity, Fig. 7a gives Jupiter's $e_p$ versus Sinope's $\Psi$ during the course of the numerical integration shown in Fig. 6b. It is clear that the region close to $\Psi=0$ is ``forbidden'' (and therefore circulation is impossible) during those times when $e_p > 0.045$. While the overall character of Siarnaq's $\Psi$ appears very different from Sinope's, Fig. 7b shows that there is a correlation between Siarnaq's $\Psi$ and Saturn's $e_p$, too. For the duration of our integration, Siarnaq's $\Psi$ follows a periodic trajectory which passes through $\Psi=\pi$ only when Saturn's eccentricity is at its minimum. This apparently happens during every third minimum in $e_p$, or about every 150,000 yrs. At all other times Siarnaq's pericenter avoids anti-alignment with Saturn's. So we conclude that the medium-term changes in the behavior of a resonant satellite's $\Psi$ appear to be forced solely by the secular variations in the eccentricities of Jupiter and Saturn. A longer-period variation has also been observed \citep{nes03} but its causes are likely to be much more complex (see below).

We can use Eq. \ref{resonant} to investigate the stability of apsidal alignment versus anti-alignment for different orbits. Since $b_1 (e)$ is always negative (Eq. \ref{yokoyamab1}), the sign of $R'_r$ depends only on $S (\cos i)$ and $\cos \Psi$. To determine which point will be a center of oscillations in $\varpi$, we need to obtain an equation of motion for $\varpi$ near the resonance. In secular resonance, $\dot{\varpi}$ is dominated by the quadrupole term, while the long-term $\dot{e}$ is caused solely by the octupole term \citep{lee03}. Since the precession of $\varpi$ caused by the quadrupole term has a zero close to the exact resonance, we can linearize $\dot{\varpi}(e)$ around that point:
$$\dot{\varpi}=K_{\varpi} (e-e_{res}),$$
\noindent where $e_{res}$ and $K_{\varpi}$ are constants. By differentiating this expression with respect to time, we get:
$${d^2 \varpi \over dt^2}=K_{\varpi}{de \over dt},$$
\noindent which, through Lagrange's equations \citep{dan92}, becomes
$${d^2 \varpi \over dt^2}=- K_{\varpi}{n a \sqrt{1-e^2} \over \mu e} {\cos i \over |\cos i|} {\partial R \over \partial \varpi}.$$
\noindent The sign function (i.e., $\cos i / |\cos i|$)is needed since $\varpi=\Omega+\omega$ for prograde bodies, but $\varpi=\Omega-\omega$ for retrograde ones.
The only term in the secular disturbing function that depends on $\varpi$ is the octupole, so we can write:
$${d^2 \varpi \over dt^2}=- {\cos i \over |\cos i|} K_{\varpi}{n a \sqrt{1-e^2} \over \mu e} {c_Y \over 64} \ S(\cos i) \ b_1 \sin \Psi.$$
\noindent If we consider that $\dot{\varpi}_p$ is constant, this can be re-written as:
\begin{equation}
\label{pendulum}
{d^2 \Psi \over dt^2} - Q_p {\cos i \over |\cos i|} K_{\varpi} \ S(\cos i) \sin \Psi=0, 
\end{equation}
where $Q_p$ is a positive quantity, independent of $\varpi$. Eq. \ref{pendulum} is the well known pendulum equation \citep{md99}. Depending on the direction of orbital motion and the signs of $K_\varpi$ and $S(\cos i)$, the stable center of $\Psi$ oscillations can be either 0 or $\pi$ (in the case of negative and positive product of those three terms, respectively). While $K_\varpi$ is a complex function of orbital elements and must be obtained through numerical integrations (cf. Saha \& Tremaine 1993), $S(\cos i)$ depends only on inclination. 
Fig. 8 plots how $S (\cos i)$ varies with average inclination. $S$ is positive for $i < 41^\circ$, and negative for higher inclinations. The positions of the four (near-)resonant satellites are marked: Pasiphae and Sinope (as well as Stephano) are in the region where apsidal anti-alignment for retrograde bodies requires $K_\varpi < 0$, whereas prograde Siarnaq's stable point of $\Psi=0$ requires $K_\varpi>0$ for its set of orbital elements. In the case of Sinope, our numerical simulations show that the episodes of negative circulation of $\Psi$ correlate with higher $e$, which indeed implies that $K_{\varpi} < 0$ (see Fig. 14 and its discussion in Section 6). Similarly, Fig. 17 of \citet{nes03} shows how negative circulation for Siarnaq coincides with smaller eccentricity, so $K_\varpi > 0$ for that moon, as expected from Eq. \ref{pendulum}. Therefore, Eq. \ref{pendulum} clearly explains why the libration center for Pasiphae and Sinope is diametrically opposite from Siarnaq's. Although the transition from one libration center to the other is related to the direction of the satellite's motion (through the different definitions of $\varpi$ for prograde and retrograde objects), retrograde satellites could, in principle, exist with orbits librating around $\Psi=0$. However, Fig 5 suggests that secular resonance is possible only for prograde bodies with inclinations above $\simeq 40^\circ$, and for retrograde satellites with inclinations below that value, making the product $\cos i \ S(\cos i)$ almost always negative. Barring some unexpected behavior of $K_{\varpi}(a, e, i)$, any new retrograde resonators that are found are likely to be in the ``Pasiphae-regime'' (librations around $\Psi=\pi$), while the prograde ones can be expected to be in the ``Siarnaq-regime'' (with $\Psi=0$ as a stable point). Exceptions would be resonators with inclinations of about $40^\circ$, at which all three relevant factors in Eq. \ref{pendulum} change sign, but resonant locking is unlikely to be strong for them in the first place, due to their smaller $a/a'$ ratio (see below).
 
 In order to probe the cause for the very different behaviors among the four objects mentioned above, we will need to take into account all the terms in Eq. \ref{rr}. Table 1 lists $a/a'$, the average $e$, $b_1 (e)$, the average $i$, $S(\cos i)$ and the relative strength of the resonant term for each of the four resonant moons; averages are taken over the length of the integration. The last quantity was obtained by dividing $R'_r$ for the each satellite by Pasiphae's $R'_r$. Here we assumed that the instantaneous eccentricities of Jupiter, Saturn and Uranus are equal to each other (they all vary between 0 and 0.1 over secular timescales). We see that Sinope's resonant term is of the same order of magnitude as that of Pasiphae (133\% of Pasiphae's), while those for Siarnaq and Stephano are one and two orders of magnitude smaller (19\% and 2\%, respectively). This comparison clearly shows why Stephano's pericenter cannot currently get locked into a resonance with that of Uranus, as its resonant term is much too weak. The low value for Stephano's resonant term mostly results from its small $a/a'$ ratio, although its eccentricity and inclination are also less conducive to resonance than is the case for the other three moons discussed here. In turn, the small $a/a'$ ratio comes not only from the relatively close orbit of Stephano ($m$=0.022), but also from the small size of Uranus's Hill sphere. The size of the Hill sphere (in terms of $a'$) depends on the planet's mass as $\mu'^{1/3}$, making the Hill spheres of Uranus and Neptune significantly smaller fractions of their $a'$ compared to those of Jupiter and Saturn. $m=(a/R_H)^{3/2}$ measures the size of a satellite's orbit relative to the planet's Hill sphere, so two moons that have the same $m$ but that orbit different planets can have very different $(a/a')$'s (for example, Himalia and the Moon have comparable $m$'s, but Himalia's $a/a'$ is an order of magnitude larger). 

But, if Sinope's resonant term is larger than Paisphae's, why is Sinope only occasionally librating, while Pasiphae appears to be deep within the libration region? The reason must lie in the initial conditions, which determine not only the libration's amplitude, but if librations are possible in the first place. It is tempting to conclude that Sinope's larger libration amplitude tells us something about the origin and evolution of these two moons. However, we must remember that integrations shown in Fig. 6 cover only 300,000 yrs, an insignificant fraction of the Solar System's lifetime. In order to make any inferences about the intrinsic differences between these moons' orbits, we need to examine their behavior over much longer times. \citet{nes03} have conducted numerical simulations of the known irregular satellites' orbits over $10^8$ yrs. They conclude that both Pasiphae and Sinope show intermittent resonant and near-resonant behaviors, with the switching between the two regimes occuring on $10^7$-yr timescales. For example, their Fig. 17 suggests that, in $4\times 10^7$ yrs, Sinope will be a strict librator while Pasiphae's $\Psi$ will circulate. Judging from these results, it is very likely that the resonant arguments of both Pasiphae and Sinope are chaotic on timescales shorter than the age of the Solar System, precluding us from obtaining any direct clues as to their origins from their present libration amplitudes. 

\citet{nes03} also found that Siarnaq occasionally exhibits relatively short episodes of pure libration. These episodes never last longer than $5 \times 10^5$ yrs, unlike the more stable states of Pasiphae and Sinope. This result is consistent with our estimate that Siarnaq's resonance is about an order of magnitude weaker than those for the two Jovian resonators. Additionally, Nesvorn\' y  et al. do not report any resonant behavior for Stephano's $\Psi$, which is in line with our conclusion that Stephano's resonant term is very weak (Table~1). 

Based on these results, as well as those of \citet{yok03} and \citet{nes03}, we can make some general conclusions about the viability of the secular resonance for different orbits. Resonant-locking is very unlikely for the orbits in the middle portion of the continuous lines in Fig. 5, as the $(a/a')$'s of such orbits are small and their inclinations put them close to the zero of $S(\cos i)$. Librations should be more common for the resonant satellites along the edges of Fig. 5, especially among the distant retrograde moons. Also, librating behavior is more likely to be seen among the Jovians and Saturnians than the Uranians, due to the relatively smaller Hill sphere of Uranus. Any present-day $\Psi$ librations among the Neptunians are highly unlikely, given the very low present eccentricity of Neptune. A satellite with a confirmed near-resonant orbit around Neptune (S/2002 N1 is the best candidate so far) might be a remnant of past locking, and therefore may indicate a higher primordial eccentricity of Neptune (see Sec. 6).  

\section{Effects of the Great Inequality}

Despite these benefits of secular models, they can never describe all the phenomena we see among the irregular satellites. For example, since they are averaged over the mean motions of the satellite and the planets, they are unable to register resonances that involve mean motions, which are known to be important for irregular satellite dynamics \citep{saha93, nes03}. We here report the serendipitous discovery of one such resonance, that involves the Great Inequality of Jupiter and Saturn, and that can have a suprisingly strong effect on the orbits of Saturnian irregular satellites. Recently, \citet{car04} have independently found several related resonances for Saturnian irregulars, although involving different resonant arguments from that discussed in this section.

Fig. 9a shows the evolution of eccentricity for a test particle orbiting Saturn. The initial conditions for the particle are $a=0.1$~AU, $e=0.4$, $i=37.5^\circ$, $\omega=90^\circ$, $\Omega=180^\circ$ and $M=45^\circ$, while the four giant planets start the integration with conditions identical to those at midnight, September 24, 2003 (see Sec. 3). This orbit should lie deep within a zone where satellite orbits are stable \citep{nes03}. However, as Fig. 9a clearly shows, the eccentricity of this test particle exhibits large and irregular variations, with the average $e$ growing from $0.3$ to almost $0.5$ in the course of our integration. Given that the time span of 300,000 yrs is just a small fraction of the total age of the Solar System, we have to conclude that such an object would probably be unstable on longer timescales, with a likely fate of colliding with Iapetus (such a collision becomes possible once $e > 0.75$). Integrations of a larger set of bodies indicate that the relevant parameter for this instability is the precession period of the longitude of pericenter. Fig. 10 plots a measure $\eta$ (defined below) of the eccentricity variation against the $\dot \varpi$ rate for 440 test particles near the one shown in Fig. 9a. These integrations spanned 30,000 yrs each and the range of initial conditions was: $a=0.1-0.13$ AU, $e=0.2-0.875$, $i=30^\circ-45^\circ$, with the steps being $0.01$ AU, $0.075$ and $1.5^\circ$, respectively (for all bodies $\omega=90^\circ$). The parameter $\eta$ is defined as 
\begin{equation}
\label{eta}
\eta=\sum^{10}_{j=1} {(<e>_j-<e>)^2 \over <e>^2},
\end{equation}
where 
$$<e>_j={10 \over T} \int^{j {T \over 10}}_{(j-1){T \over 10}} e dt$$
and
$$<e>={1 \over T} \int^{T}_{0} e dt.$$
A high value of $\eta$ for a test particle indicates that $e$ has either a long-period ($>$ 3000 yrs) oscillation, or a secular trend. Fig. 10 clearly shows the close correlation between the particles' $\eta$ and $\dot{\varpi}$. The high--$\eta$ particles tend to have $\dot{\varpi} \simeq 0$ or $\dot{\varpi} \simeq 5.4 \times 10^{-4}$. The first of these two features can safely be identified with the secular resonance (Sec. 4). The apsidal precession period associated with the second feature is about 1860 yrs, which is close to twice the period of the Great Inequality of Jupiter and Saturn (883 yrs). The Great Inequality is a consequence of the mean motions of Jupiter and Saturn being close to, but not in, 5:2 commensurability. This near-resonant perturbation has a significant effect on the orbits of both planets, and it prevented many early reasearchers from completely harmonizing theories based on Newtonian gravity with the observed planetary motions \citep{baum97}. The 883-year period corresponds to the residual motion of the near-resonant argument $5 \lambda_S - 2 \lambda_J$. In order to make a valid term in the disturbing function, consistent with d'Alembert's rule (which requires the sum of numerical coefficients to vanish), this argument has to be complemented by at least three other angles, with negative coefficients. The longitudes of the lines of apsides and nodes of Jupiter and Saturn (which change slowly) commonly fill this role, so a number of observed strong perturbations of these planets' orbits have periods close to 900 yrs. Based on all this, it is straightforward to conclude that if we subtract twice the $\dot{\varpi}$ frequency of our high-$\eta$ test particles, the resulting argument would change exceptionally slowly. Since half of the apsidal precession period of those test particles is still longer than the 883-year Great Inequality period, we conclude that the third additional term should be a prograde secular angle, with $\varpi_S$ being the obvious choice. So we define our candidate resonant angle $\xi$ as
\begin{equation}
\label{xi} 
\xi=5 \lambda_S - 2 \lambda_J - 2 \varpi - \varpi_S.
\end{equation}
Fig. 9b plots the evolution of $\xi$ for the test particle whose eccentricity history was shown in Fig. 9a. Not only does $\xi$ librate at times, but the correlation between the behaviors of $e$ and $\xi$ is impressive. Apparently, the largest change in the eccentricity occurs when $\xi$ librates, whereas $e$ has much less of a secular trend at times when $\xi$ circulates. On the time scale of this integration, the evolution of $\xi$ appears to be chaotic, although there yet may be some longer-term periodicities. In any case, we feel that the identification of this perturbation with the Great Inequality, and the specific resonant argument $\xi$ is firm. We should also note that Fig. 10 suggests that no other harmonics of the Great Inequality appear to affect the test particles appreciably, at least within the range of apsidal precession frequencies found among our set of test particles (cf. Carruba et al. 2004). 

We now address the circumstances (combination of orbital elements) for which the Great Inequality resonance occurs. Formally, it should happen for any bodies orbiting Jupiter and Saturn whose $\varpi$ circulation periods are around 1800 yrs. We found that, while detectable, this resonance does not strongly perturb Jovian satellites (even if $\varpi_J$ is substituted in place of $\varpi_S$ in Eq. \ref{xi}), or those retrograde Saturnians for which relation (\ref{xi}) holds (this is possible for some orbits with inclinations slightly above the secular resonance). Effects as large as those displayed in Fig. 9a are restricted to the prograde Saturnians. Saturn's satellites are likely favored due to the larger effects of the Great Inequality on that planet's orbit, but the reason for the prograde bias is likely to be more complex. Fig. 11 locates the Great-Inequality resonance in $m-i_{min}$ space. The top three continuous lines plot the resonant location for different eccentricities (see caption). It is not surprising that no satellites at the present time inhabit the resonant region (the proximity of Siarnaq to the top line is irrelevant, as that line applies to bodies of much lower eccentricity). It is even more interesting to explore how the location of the resonance shifts with the changing period of the Great Inequality. Since the latter measures the distance of Jupiter and Saturn from the exact 5:2 resonance, even relatively small changes in the orbital periods of those two planets can lead to a large variation in the Great-Inequality's period. Since Saturn probably migrated outward and Jupiter inward in the early Solar System \citep{hahn99}, they should have been further from the resonance in the past (since $5 n_S - 2 n_J > 0$). The bottom solid line in Fig. 11 plots the location of the Great Inequality resonance for the hypothetical epoch when $(5 \dot{\lambda}_S - 2 \dot{\lambda}_J)^{-1} = 500$ yrs (we chose $e_{max}=0.6$, which is a good approximation for most prograde Saturnian irregulars). With this change of parameters, the resonance moves to signficantly lower inclinations, approaching the elements of Albiorix, which presently has $P_\varpi \simeq 820$ yrs; so a somewhat shorter period of Great Inequality ($\simeq 400$ yrs) would definitely affect it. Such a shift in the Great-Inequality's period requires a change in Saturn's period of less than 1\%, and $\Delta a_S < 0.1$ AU. Migration on that, or larger, scale is likely for a wide range of initial parameters of the primordial planetesimal disk \citep{hahn99}. 

Since Albiorix is the largest member of a satellite cluster that also includes Erriapo and Tarvos (Gladman et al. 2001; we will call it the ``Gaulish cluster''), the history of this resonance may be coupled to the origin of the whole group. It is unclear whether this is a collisional group \citep{nes03} and, if it is, if the passage through the Great-Inequality resonance happened before or after the collisional break-up. \citet{nes03} find that the velocity dispersion of the cluster is too large (30--60 m/s) to be explained solely by the velocity distribution of fragments from a catastrophic collison, given that the parent body's diameter could hardly be larger than 50 km. They conclude that, if the Gaulish cluster is indeed collisional, some other process must have dispersed the fragments after the original disruption. It is tempting to suggest that sweeping by the Great-Inequality resonance could have additionally dispersed this group, but we do not think that such a scenario is likely, at least in its simplest form. The semimajor axes of the Gaulish cluster members are much further apart in velocity terms than their $e$ and $i$. The Great-Inequality resonance does not affect $a$ significantly, but rather tends to induce wide variations in $e$. In our opinion, the present distribution of the members of the Gaulish cluster could be better explained through a {\it depletion} by the Great-Inequality resonance, rather than by dispersion. We suggest that the Gaulish cluster may have originated from the disruption of a body much larger than 50 km in diameter. Subsequent sweeping by the Great-Inequality resonance through the Gaulish cluster eliminated much of its material. In this view, the three known satellites were among those that managed to survive this process, by leaving the resonance before they were lost to escape or to collisions with major moons. The erratic nature of any planetary migration would offer many possibilities for escape and capture events. The very similar $\varpi$ periods of Erriapo and Tarvos (682 and 674 yrs, respectively) indicate that they would have escaped the resonance during the same epoch.

Direct numerical simulation of the possible effects of planetary migration on the Gaulish cluster is clearly needed. In a recent study, \citet{nes04} suggest that the high flux of impactors in the early Solar System probably dominated the collisional history of many irregular moons. Therefore, it is not completely unexpected that the disruption of the parent body might have preceded much of the planetary migration. It is possible that a detailed study of the Gaulish cluster's history could directly constrain the timing of the breakup event (if there were one) in relation to the planetary migration. However, while the Great-Inequality resonance is very interesting dynamically, its significance is limited to just a few of the Saturnian irregulars. At this point, the only conclusion  we can make about the history of the Gaulish cluster is that its present characteristics are most likely not primordial, but modified by the Great-Inequality resonance. Because of all this, it would be risky to infer the origin of the cluster's progenitor based on the present parameters of its members.

\section{Implications for the Origin of Irregular Moons}

Most researchers agree that the irregular satellites are captured bodies, which formed in the protosolar nebula independently of the planet. Several capture mechanisms have been proposed; the most prominent ones are collisions \citep{col71}, increase in planetary mass \citep{hep77, vn04} and aerodynamic drag \citep{pol79}. 

Various aspects of the possible capture and subsequent orbital evolution of Jupiter's largest irregular satellite, Himalia, starting from the gas-drag hypothesis, have been explored by \citet{cuk04}. They find that the capture of Himalia by aerodynamic drag is possible, although their model for the Jovian nebula differs from that of \citet{pol79}. Himalia is arguably the logical starting point if one is interested in exploring the origin of the irregular satellites. Since it is both relatively large and prograde, it required more dissipation to be permanently captured than any other irregular (with the possible exception of Triton, whose origin is outside the scope of this work). If Himalia were captured by dissipation in the nebula, such an origin would be even more likely for other irregulars. A more subtle difference between Himalia and most other irregulars is that Himalia appears to be outside of the region where resonances are common. \citet{cuk04} find that this was likely true in the past, since no detectable resonant events can be seen during numerical simulations of its post-capture orbital evolution. Using results from previous sections, we can confirm this result and put it in the wider context of irregular-satellite dynamics. Fig. 12 is essentially the same as Fig. 5, only that now all satellite groups are labeled. The clusters containing bodies with diameters larger than 100 km have their names marked with asterisks. It is clear that Himalia (in the middle right) lies some distance away from the secular resonance. Not only is this true at the present epoch, but \citet{cuk04} argue that the post-capture elements of Himalia were probably close to $m=0.13, i=40^\circ$, which is also comfortably below the secular resonance. Therefore the result that Himalia's orbital evolution avoided resonant passages should not be surprising. 

Globally, Fig. 12 shows significant clustering of the irregular satellite groups around the locations where the secular resonance is possible. We will refer to this super-family as the ``Main Sequence'', due to its superficial visual similarity to the famous feature on the Hertzsprung-Russell diagram. This  grouping naturally includes all objects that are in (or very close to) secular resonance, including Pasiphae (labelled Ps in Fig. 12), Sinope (Sn), Siarnaq (Sr), Stephano (St), possibly S/2003~S1 and S/2003~J2 (off scale, at $m=0.225$), as well as some small Jovian satellites that are likely collisional fragments of Pasiphae and Sinope. Almost all Kozai librators are on the ``Main Sequence'', too, since the Kozai resonance's boundary (Figs. 4a and 4b) lies usually only several degrees above the secular resonance, and the known librators are mostly found just above the boundary. These include Kiviuq and Ijiraq (K), Euporie (E), S/2003~J20, S/2003~U3 and possibly S/2002 N2 and N4. Another dynamical class of objects comprising the ``Main Sequence'' are reverse-circulators. These are the objects whose orbital elements put them between the secular and Kozai resonances. Their arguments of pericenter are circulating, but more slowly than their nodes. This leads to the precession of $\varpi$ being dominated by $\dot \Omega$, which is always in the opposite direction from the orbital motion (Eq. \ref{kozaio}). The satellites exhibiting reverse-circulating behavior are Ananke with its family (A), Themisto (T), Paaliaq (Pl), Caliban (Cb), and, according to the current orbital solutions, S/2002 N1 and S/2003 N1. Finally, several objects seem to lie relatively close to the secular resonance but still have their $\varpi$ circulating regularly in the direction of their orbital motion. At this point, only retrograde objects are known to do this, and they include Carme with its cluster, Skadi and probably S/2001~U3. Here we need to caution that the orbits of some of the objects we listed above as prospective members of the ``Main Sequence'', especially the new Neptunian satellites, are still somewhat uncertain. However, we believe that it is highly unlikely that there is a systematic error in the present orbit solutions that can make them appear closer to the resonances than they really are. If anything, a preliminary orbit is likely to miss the resonance. The correction to the orbit of S/2003 S1 based on observations early in 2004 (Sheppard {\it et al.} 2004) puts it closer to the secular resonance than the solution plotted in Figs. 5 and 12. 

Table 2 lists the largest members of each suspected irregular satellite cluster, and recapitulates our classification of them into dynamical groups. For all objects, the $\nu=-\dot{\varpi}/\dot{\Omega}$ ratio is also given, in order to demonstrate that our classification has a direct quantitative basis.  The definition $\nu$ was chosen so that in the ideal case of $m<<1, e<<1$ and $\sin i<<1$, $\nu=1$. It can be higher for low-$i$ prograde moons (e.g., for the Moon, $\nu=2.09$), while it is generally lower for inclined prograde, and all retrograde, orbits. It is obvious that $\nu=-1$ for Kozai resonance and $\nu=0$ for the secular resonance. Reverse-circulators have negative $\nu$'s (usually close to zero, though) while the objects with $\varpi$ precessing in the direction of orbital motion have $\nu>1$. Among the latter objects, the distinction between those that are, and are not, close to the ``Main Sequence'' is not sharply visible in their $\nu$. However, it is uniquely defined, with objects having $\nu<0.2$ appearing close to the ``Main Sequence'' in Fig. 12. On the other hand, the classification of irregulars into clusters and the choice as to which objects are more-or-less primordial are open to debate \citep{nes03, grav03}. We decided to be conservative about cluster affiliation: all the irregulars whose cluster membership is not clear are taken to be independent objects (i.e., we chose to ``split'' clusters in ambiguous cases). This principle is somewhat softened for the Uranian irregulars, where we identify only four ``clusters'': S/2001 U3 and S/2003 U3 individually, Caliban and Stephano together, and all other objects belong to Sycorax's family. Notice that Sycorax's escape velocity (which determines the dispersion of fragments) is a much higher fraction of its orbital speed than the same ratio for any other irregular, making a very extended cluster not too surprising. The extent of the Sycorax cluster was likely increased by post-breakup gravitational scattering among family members \citep{chr04}. In the case of Neptune, by assigning each satellite to its own group, we have most likely overestimated the satellite diversity, and it is very probable that the two pairs with similar mean motions might end up being genetically related. However, at this time, we believe that such a conclusion would be premature.

It is interesting to note that none of the three more massive objects (Himalia, Phoebe and Sycorax) have orbits anywhere close to the secular resonance. While it is hard to claim any statistical significance with only three points, we aver that this dichotomy does imply some kind of significant difference that sets the largest few irregulars apart from all others. This dichotomy is not surprising because the aerodynamic accelerations -- vital in satellite capture -- act differently on different-sized bodies. In our opinion, large bodies were likely captured before the rest, at an epoch when the gas density was high enough to cause permanent capture of a 200-km planetesimal. \citet{cuk04} find no mechanism that can arrest the decay of proto-Himalia into Jupiter except for a fast ($10^4$-yr timescale) evolution of the nebula itself. Such a capture scenario would not result in any preferential final orbit for the satellite. 

On the other hand, smaller satellites could not have been present at the same epoch, since the strong gas drag would make them rapidly spiral into Jupiter. Therefore, we think that the capture of most of the members of the ``Main Sequence'' postdates that of Himalia, Phoebe and Sycorax. The fact that a large fraction of the satellite groups are found  close to the secular resonance indicates that some kind of discriminating process must have acted on these objects, causing them to end their evolution as members of the ``Main Sequence''. The large amount of orbital evolution that some of these objects experienced (Themisto, for example, is very tightly bound) and the existence of several objects that are exactly in the secular resonance, suggests that gas drag played a role in the capture of the smaller irregulars, too. Their clustering in the region of slow $\varpi$ precession hints that some process acting exclusively on slowly-precessing orbits also had a role perhaps in their capture and certainly in their evolution.

Two schemes might explain the ``Main Sequence'': it might be a region where the satellites halted their orbital decay into the planet, which is otherwise an unavoidable consequence of a satellite capture in a long-lived gas disk. On the other hand, the ``Main Sequence'' might result from a bias: permanent capture might have been possible but rare at some time in the past, with the planetesimals being captured only into certain orbits, which led to the clustering of resulting irregular satellites. We are unable unambiguously to endorse either of these hypotheses. Nonetheless we discuss each hypothesis briefly since we have tried (inconclusively) to confirm some of their predictions. 

The simplest way of producing the ``Main Sequence'' would be if the resonant interaction supplied energy to objects that otherwise would have decaying orbits. However, such a mechanism has to be rejected on theoretical grounds. The argument of secular resonance has no dependence on the satellite's mean longitude (see Sec. 4 below Eq. \ref{resonant}; by definition, secular terms cannot include $\lambda$ which varies rapidly). Since the mean longitude (or the anomaly, depending on one's choice of Hamiltonian variables) is the conjugate variable to the total energy of an orbiting object \citep{md99}, only a term containing the instantaneous orbital position can be expected to induce changes to an orbit's semimajor axis (which is equivalent to the total energy). We are not aware of any published numerical experiments that support the contrary conclusion. We also recall here that our preliminary numerical experiments on gas-drag evolution (cf. \' Cuk \& Burns 2004) with secular resonances never found truly permanent capture.

Another, subtler way that passage through the secular resonance can stop orbital decay is by changing the body's eccentricity. The survival of retrograde satelites, especially Phoebe with its low eccentricity orbit, strongly suggests that the circumplanetary disk must have had a sharp drop in surface density, most likely near Phoebe's present orbit (for details, see \' Cuk \& Burns 2004). In this view, just the pericenter of most satellites penetrated the nebula so that aerodynamic drag affected them only during close approaches. Therefore, a sharp change in the eccentricity of the satellite (which would in turn modify the pericenter distance) could radically alter the rate of satellite's evolution. If a satellite experienced a significant decrease in eccentricity due to secular-resonance passage, it could theoretically become decoupled from the disk altogether and end its orbital evolution in the resonance (or very close to it). Since Sinope is not only affected by the strongest resonant Hamiltonian (Table 1), but also exhibits passages from libration to circulation, we will use this irregular as a test-case for this hypothesis.  Saha \& Tremaine (1993, Fig. 2) have previously described this transition between circulation and resonance, and show the results of several integrations from $10^5$ yr to $2 \times 10^6$ yr with all giant planets and with Jupiter alone. Fig. 13a shows the behavior of Sinope's eccentricity and its resonant argument (during the simulation shown in Fig. 6b) in polar coordinates, i.e., $e$ is shown by the distance from the center, while $\Psi$ is represented by the angle from the x-axis. Clearly the allowed positions of ($e, \Psi$) in this plot lie on two rings that overlap at $\Psi=0$. Motion around the rings corresponds to the two directions of $\Psi$'s circulation, while the librations are equivalent to crescent-shaped arcs in which the particle shifts from one ring to the other without crossing the positive x-axis. Passage through the resonance would be seen in this plot as switching from circulation along one ring to the opposite motion along the other one. To identify the direction of $\Psi$'s circulation along the rings, Fig. 13b plots how $e$ changes with time during the same simulation. A comparison of Figs. 6b and 13b reveals that the periods of retrograde motion of $\Psi$ are correlated with high eccentricity, therefore on the outer ring $\dot \Psi<0$ while $\dot \Psi >0$ on the inner ring. We know that orbits with inclinations below that required for the secular resonance have $\varpi$ circulating in the orbital direction, while the higher-inclination ones are reverse-circulators. However, since the secular resonance appears to exhibit a downward turn as $m$ increases for retrograde orbits (Fig. 5), the reverse-circulators ($\dot \varpi>0$) are to Sinope's left and objects with $\dot \varpi<0$ are to the right. Hence, if an object is evolving due to gas drag from larger to smaller $m$ (i.e., $\dot a <0$), it would increase its eccentricity (and correspondingly accelerate its orbital evolution) as it crosses the secular resonance. Therefore we do not believe that this mechanism could have arrested the orbital decay of the retrograde satellites. 

For prograde bodies, the shape of the secular resonance (Fig. 5) suggests that a decaying satellite would cross the resonance from the region of direct precession to that of $\varpi$'s reverse circulation. Fig. 19 in \citet{nes03} shows that the instances of retrograde circulation of Siarnaq's $\Psi$ are correlated with an eccentricity lower than that seen during libration. Therefore, among the prograde satellites, $\Delta e$ during the resonance passage would have been negative, unlike for distant retrograde satellites. However, Fig. 19 of \citet{nes03} also demonstrates that this change is rather small for Siarnaq, making any impact of the resonant passage much less dramatic than it would be for a Sinope-type orbit. With this in mind, we conclude that the change in $e$ during the passage through the secular resonance does not make a very promising candidate for a process leading to the clustering of irregular satellites into the ``Main Sequence''.

The opposite view, that the ``Main Sequence'' might mirror a bias in the capture mechanism, has some support in the findings of \citet{cuk04}. Their Fig. 7 shows that the timescale for temporary capture is related to the precession period of $\varpi-\varpi_{planet}$. This is because capture preferentially happens when the planet is at pericenter (and zero velocity curves open up the most; Murray \& Dermott 1999, Hamilton \& Burns 1992), accordingly it results in a temporary satellite orbit having its apocenter close to the Lagrangian point. It is easy to see that the combination of these two constraints requires $\varpi-\varpi_{planet}=0$ at the capture. For the duration of temporary capture, the planetesimal experiences resistance from the circumplanetary disk during each pericenter passage, and loses orbital energy and momentum. If there could be some preferred orbits where temporary capture could last significantly longer than average (usually it persists for about 100 years in the case of Jupiter; see the arguments of \' Cuk \& Burns 2004), such objects could become captured even if the nebular density were relatively low. In such a nebula, any subsequent collapse of satellite orbits due to continuing gas drag would be much slower, and the chance of them surviving to the present day would be much higher.

According to Fig. 12, the region of secular resonances widens towards larger $m$'s for both prograde and retrograde orbits, actually becoming rather large for very distant retrograde objects. At very large $m$, satellite orbits become unstable. From the considerations outlined immediately above, we would expect that locales in the orbital-element space where the ``Main Sequence'' reaches the stability limit for distant orbits will be places where temporary capture is most likely to become permanent even in the presence of only moderate gas drag. In this scenario, a chaotically changing orbit of a temporarily captured object could  drift into one of  these ``keyholes'' at the ends of the ``Main Sequence''. A temporary-satellite orbit whose orbital elements put it close to the ``Main Sequence'' would have a relatively slow precession of $\varpi$, and this would delay its escape, since $\varpi-\varpi_{planet}=0$ must be satisfied for an orbit to slip out of the Hill sphere's opening. This delay might prolong the temporary capture phase by more than an order of magnitude (Table 2), as $|\nu|<0.1$ is typical for objects close to ``the Main Sequence'' (but not in Kozai resonance) while $\nu \simeq 1$ for other satellites (Table 2). After the planetesimal has been permanently captured, its orbital evolution would decrease its $a$ and $e$, while the inclination would lessen for prograde orbits and rise up (i.e., move away from $i=180^\circ$ and toward $i=90^\circ$; \' Cuk \& Burns 2004) for retrograde ones. Such a migration would force a prograde satellite toward the left and down in Fig. 12, while a retrograde one would move to the right and up. This way, the bodies that start at the ends of the ``Main Sequence'' would stay close to it even after substantial orbital migration. In such a scheme, all the moons would be captured into orbits with circulating $\omega$ but they could subsequently evolve into orbits with librating $\omega$. Finally, today's extant objects in the secular resonance might have been captured into it later either by very weak residual gas drag, sweeping of resonances due to planetary migration \citep{nes03}, small growth in the planet's mass, or even a variant of the Yarkovsky effect \citep{cuk02b}. A large population of bodies in the general vicinity of the secular resonance would make such a capture likely, even if the changes in the position of the satellites and the resonance relative to each other in the orbital-element space was not dramatic.

Studies of the short-term stability of distant asteroidal satellites by \citet{ham91} appear to support the ``keyhole'' hypothesis. They integrate numerous test particles starting on circular orbits that exhibit a range of $a$ and $i$. Their Fig. 15 illustrates the initial semimajor axes and inclinations of the particles that survive in orbit for five asteroid years (equivalent to about 60 years for the comparable problem with Jupiter). This plot clearly shows that there are two inclination values for which stable orbital distances reach local maxima: 50$^\circ$--60$^\circ$ ($a_{max}\simeq 1/2 R_H$) and 160$^\circ$--180$^\circ$ ($a_{max}\simeq R_H$). We point out that, in the context of the Kozai cycle, these are to be considered maximum inclinations, since they coincide with minimum eccentricities. Having this in mind, these two distant stability regions are clearly correlated in inclination with our proposed ``keyholes''. The critical distances found by \citet{ham91} are much larger than those corresponding to long-term stability, but it is likely that the dependence of critical distance on $i$ should be somewhat similar. More recently, \citet{vn01} have studied the stability of temporarily captured satellites of Uranus, and their results lend further support to our ``keyhole'' hypothesis. \citet{vn01} numerically followed orbits of arbitrary eccentricity and inclination, and found a dependence of the capture duration on inclination (their Fig. 4b) that is very similar to the results of \citet{ham91}.  Although dealing with orbits around objects of very different masses, these numerical experiments show that the stability of distant orbits clearly depends on their inclination, and it is likely that this dependence will persist for the orbits of permanently captured objects. 

At the present time, we favor the ``keyhole'' hypothesis as the explanation for the clustering of small irregular satellites around the secular resonance. Only large-scale direct numerical simulations of the capture process could help us decide between the different scenarios. Since we have every reason to think that the likelihood of satellite capture and subsequent survival is low, many thousands, if not millions, of particles might need to be followed before a statistically significant sample of hypothetical irregular satellites could be acquired. The uncertainty in the details of both planetary formation and migration would require multiple simulations in order to explore different choices for uncertain parameters. Finally, once such hypothetical irregular satellite systems are generated, bodies on inclined orbits that are unstable on long timescales \citep{car02} would need to be eliminated, since their presence would make it harder to compare the synthetic satellite system to the natural ones \citep{ast03}. Such a project requires substantial numerical resources and lies outside this paper's scope, which deals primarily with an analytical description of today's irregular satellite orbits. 

Apart from the large irregulars, Saturn's ``Gaulish'' (Albiorix, Erriapo and Tavros) and ``Norse'' (Mundilfari, Suttung, Thrym and Ymir) clusters are also not in the region adjacent to the secular resonance. In Sec. 5 we showed that the Gaulish cluster was likely affected in the past by the Great-Inequality resonance, which is specific to Saturn, and therefore cannot be considered to indicate the primordial irregular-satellite population. We suspect that some related resonance might have also affected the Norse cluster, which is not only more dispersed in terms of their relative velocities that one would expect for a collisional family, but also exhibits a baffling distribution of eccentricities, which vary more than $a$ and $i$ among the cluster members. Clearly, more research is needed to better appreciate how planetary migration and the resulting shift in solar-system secular frequencies affect irregular satellites, especially those of Saturn (cf. Carruba et al. 2004). One should also recognize that other resonances of a similar type, involving either the Great Inequality or perhaps the ``Lesser Inequality'' of Uranus and Neptune (associated with $2 \lambda_N-\lambda_U$ argument), could exist at other planets. Alternatively, satellite-satellite scatterings might also be important \citep{chr04}, especially for members of the ``Norse'' cluster that share their orbital space with Phoebe.       

\section{Summary}

In the previous sections, we have shown how a mixed analytical-numerical approach can greatly improve our understanding of the irregular-satellite dynamics. Of course, plots similar to Figs. 4b and 5 can as well be generated simply by direct numerical integration, and doubtless such plots would be more faithful to the behaviors of the real satellites. However, it is noteworthy that the generation of the grid of test particles on which Figs. 4b and 5 are based, using our secular model, took only about an hour of computing time (on an ordinary personal computer with a 2-GHz processor). The same calculation using a symplectic integrator would require either several weeks of computing time or a much more powerful processor. While such a project would be by no means exceptional, we maintain that a reasonably accurate approximation always has certain benefits over a ``brute-force'' approach. Seeing how the Kozai, evection and octupole terms in the disturbing function interact and produce the observed distribution of resonances in the orbital-element space provides deeper insight into the irregular-satellite dynamics than we would have just by finding the resonant locations numerically.

The dynamics of the irregular satellites has proven to be both exceptionally rich and difficult, stimulating a steady flow of publications from several research groups during the last few years. This paper deals with many areas of irregular-satellite research; its main conclusions are:

1. The evection and other short-period terms, which do not vanish after averaging over the planetary orbital motion, must be included in the disturbing function when constructing an accurate secular theory of irregular-satellite motion.

2. Our model has purely secular and evection terms derived directly, while the total contribution caused by other short-period terms is synthesized on the basis of classical lunar theories \citep{tis94, bro96}. This model is found to predict rather accurately the precession rate of the longitude of pericenter ($\varpi$) for almost all orbital elements. The locations of secular resonances can be calculated correctly using this model, while the critical inclination for Kozai resonance can be computed for all but very eccentric, retrograde orbits (Figs. 4 and 5).

3. By analyzing the octupole term in the disturbing function \citep{yok03}, we show that both the locking strength and the location of the stable point of the secular resonance can be estimated from relatively simple theoretical considerations, and that these predictions agree well with direct numerical integrations of the known irregular satellites' motions.

4. We report a serendipitous discovery of a new secular resonance in the Saturnian system, which consists of a 1:2 commensurability between a satellite's apsidal precession frequency with the Great Inequality of Jupiter and Saturn. This resonance is very strong and can make the orbits of satellites unstable on short timescales; its location depends sensitively on the present orbits of Jupiter and Saturn, and it likely swept through the ``Gaulish'' cluster of Saturn's irregular satellites during planetary migration (cf. Carruba et al. 2004).

5. We observe that the large majority of irregular-satellite clusters inhabit the region close to the secular resonance. We refer to this assemblage as the ``Main Sequence''. Only the largest satellites ($R>100$ km) and those Saturnians likely affected by the Great-Inequality resonance (see preceding point) clearly do not belong to the ``Main Sequence''.

6. We propose a new variant of capture under aerodynamic drag as the mechanism for the formation of the ``Main Sequence''. \citet{cuk04} have shown that satellite capture and escape require an approximate alignment of the apses of planet's and the satellite's orbits. Based on this, we argue that temporary capture should last longer for any orbits having slow apsidal precession, making them more likely candidates for permanent capture.

\acknowledgments

The authors wish to thank Valerio Carruba, Bob Jacobson, David Nesvorn\' y and Phil Nicholson for their help on our irregular satellite dynamics project over the last three years. We also thank the anonymous reviewer and especially PDN for comments on an earlier draft.

\newpage

\begin{figure}
\label{empirical}
\includegraphics[angle=270, scale=.60]{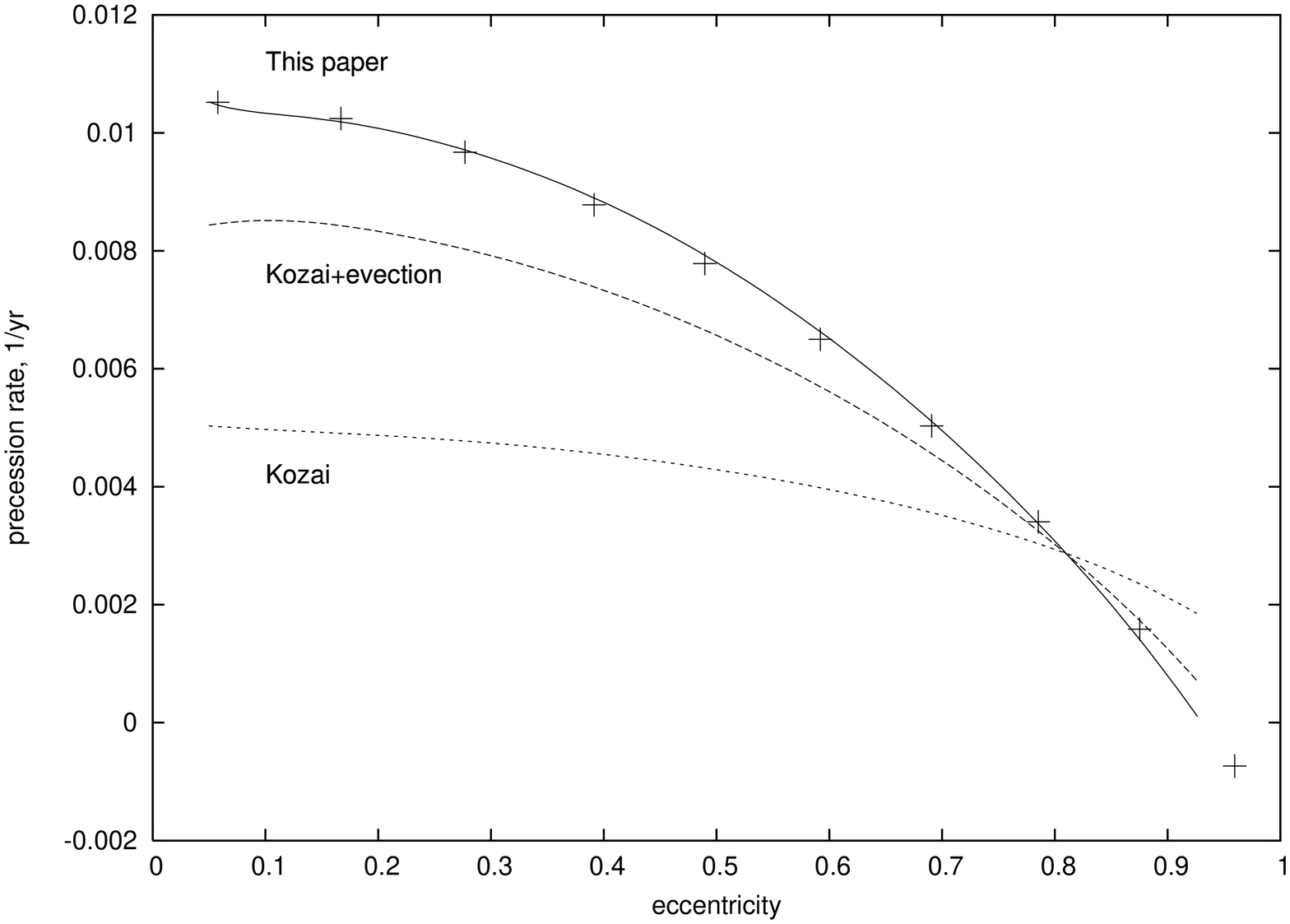}  
\caption{The dependence of the apsidal precession rate $\dot \varpi$ on satellite's eccentricity, for a near-planar prograde satellite ($m=0.08, i=5^{\circ}$) moving about a low-eccentricity pseudo-Jupiter ($P'=11.86$ yr, but $e'=0.01$). Crosses plot the result of a symplectic integration, while lines plot the predictions of secular theories. The precession rates for the latter were obtained by advancing the secular equations of motion using a  Burlisch-Stoer numerical integrator. The short-dashed line plots the predictions of Kozai's theory, the long-dashed curve also includes the evection terms, while the solid line depicts the results of the mixed model derived in this paper (Eqs. \ref{finaldwdt} and \ref{finaldodt}).}
\end{figure}

\newpage
\newcounter{subfigure}
\renewcommand{\thefigure}{\arabic{figure}\alph{subfigure}}
\setcounter{subfigure}{1}

\begin{figure}
\label{demoep}
\includegraphics[angle=270, scale=.5]{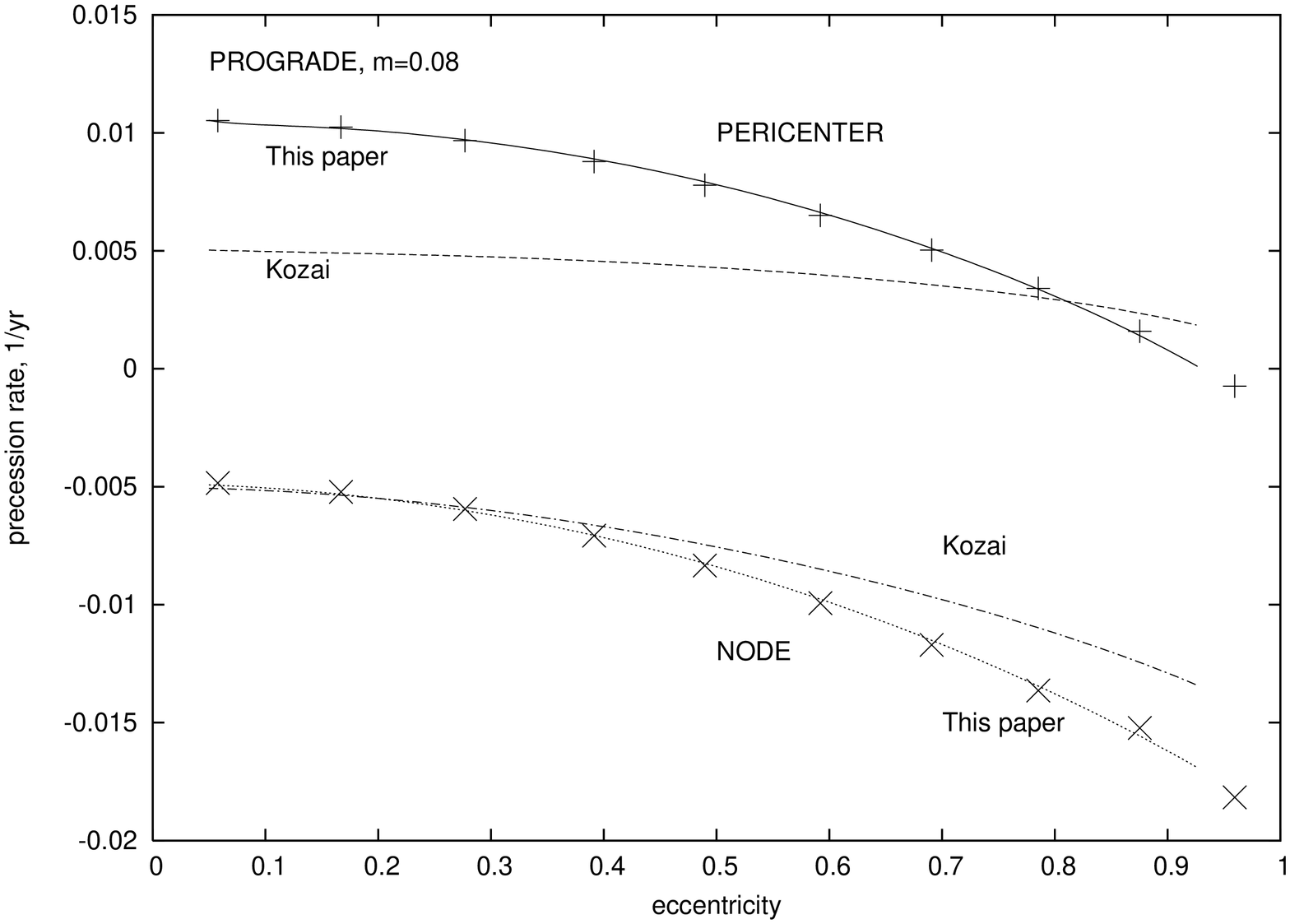}  
\caption{}
\end{figure}

\addtocounter{figure}{-1}
\addtocounter{subfigure}{1}
\begin{figure}
\label{demoeq}
\includegraphics[angle=270, scale=.5]{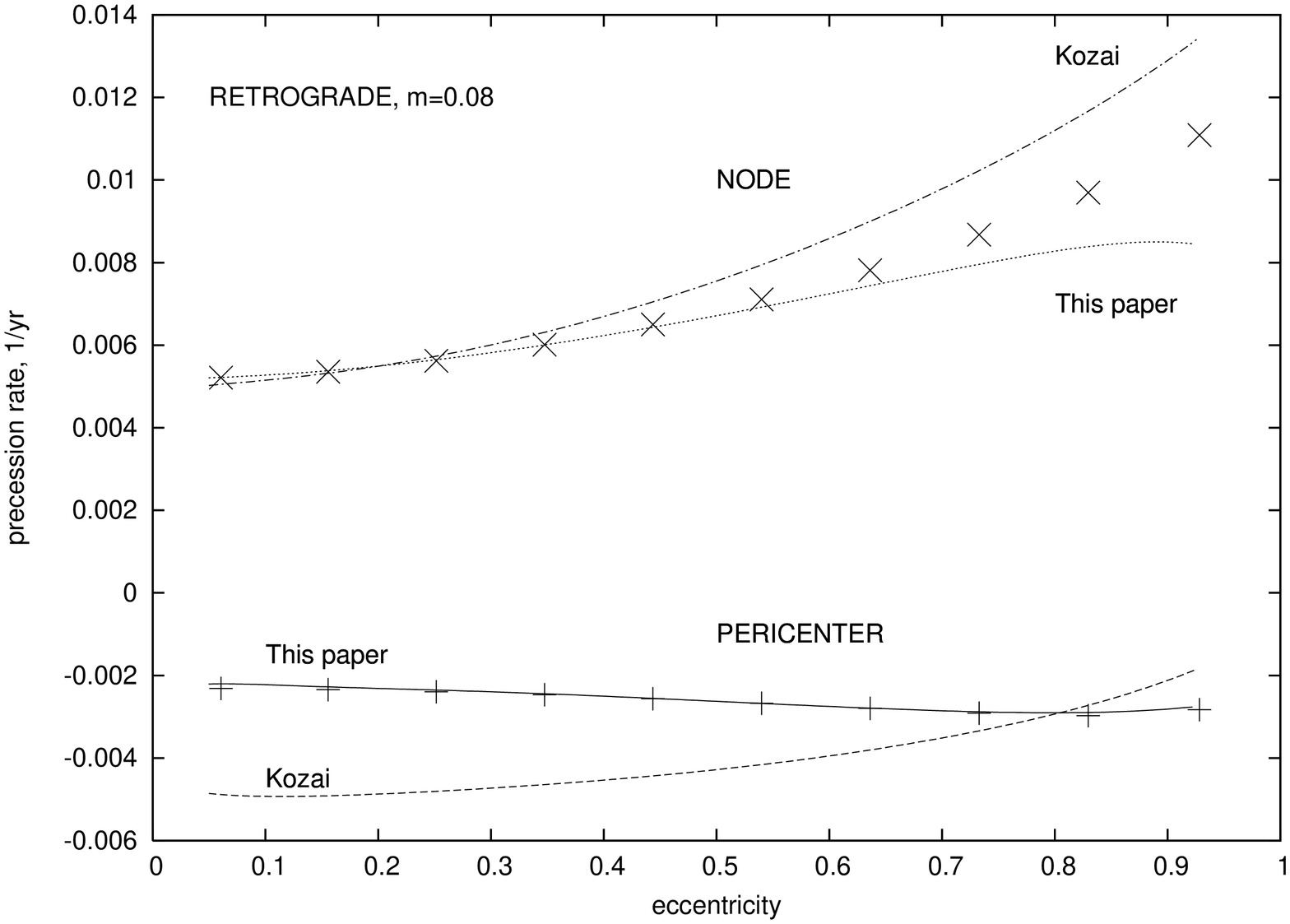}  
\caption{}
\end{figure}

\addtocounter{figure}{-1}
\addtocounter{subfigure}{1}
\begin{figure}
\label{demoer}
\includegraphics[angle=270, scale=.5]{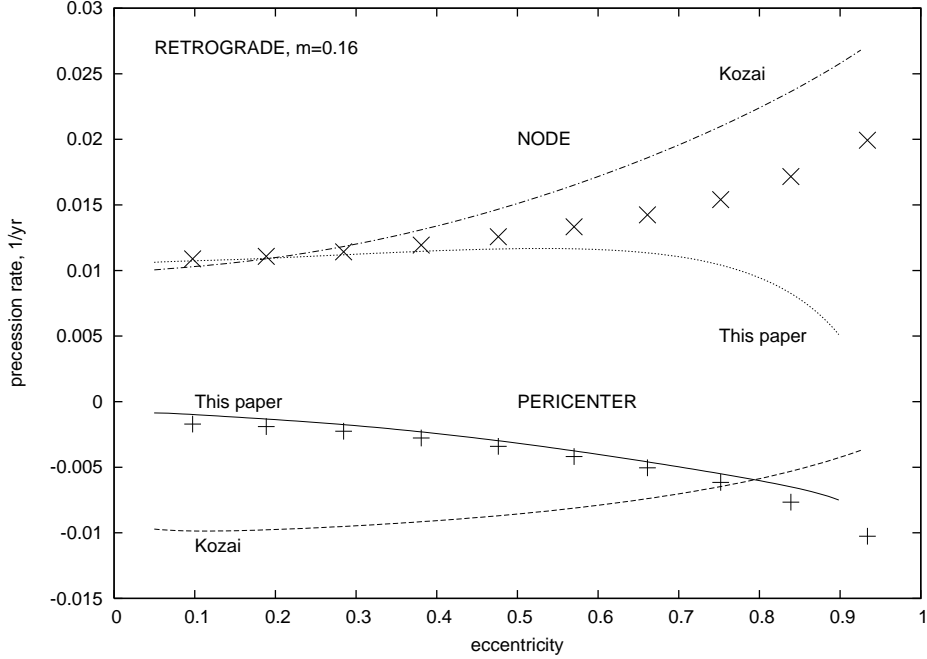}  
\caption{Similar to Fig. 1, for various eccentricities, comparing the motions of both the line of apsides and the line of nodes predicted by secular theories with direct numerical integrations. In all three panels, the precession rate $\dot \varpi$ for the line of apsides predicted by Eq. \ref{finaldwdt} is plotted as a solid line; likewise, the predicted precession rate $\dot \Omega$ for the line of nodes (Eq. \ref{finaldodt}) is plotted by a dotted line. The same two quantities, predicted on the basis of the Kozai theory (Eqs. 4 and 5) are plotted by dashed and dashed-dotted lines, respectively. All orbits shown have low inclination ($i=5^\circ$) and scaled orbital periods of a) $m=0.08$ (prograde), b) $m=0.08$ (retrograde), and c) $m=0.16$ (retrograde). The planet's period is taken to be that of Jupiter (11.86 yrs).}
\end{figure}

\newpage

\renewcommand{\thefigure}{\arabic{figure}\alph{subfigure}}
\setcounter{subfigure}{1}
\begin{figure}
\label{demoip}
\includegraphics[angle=270, scale=.5]{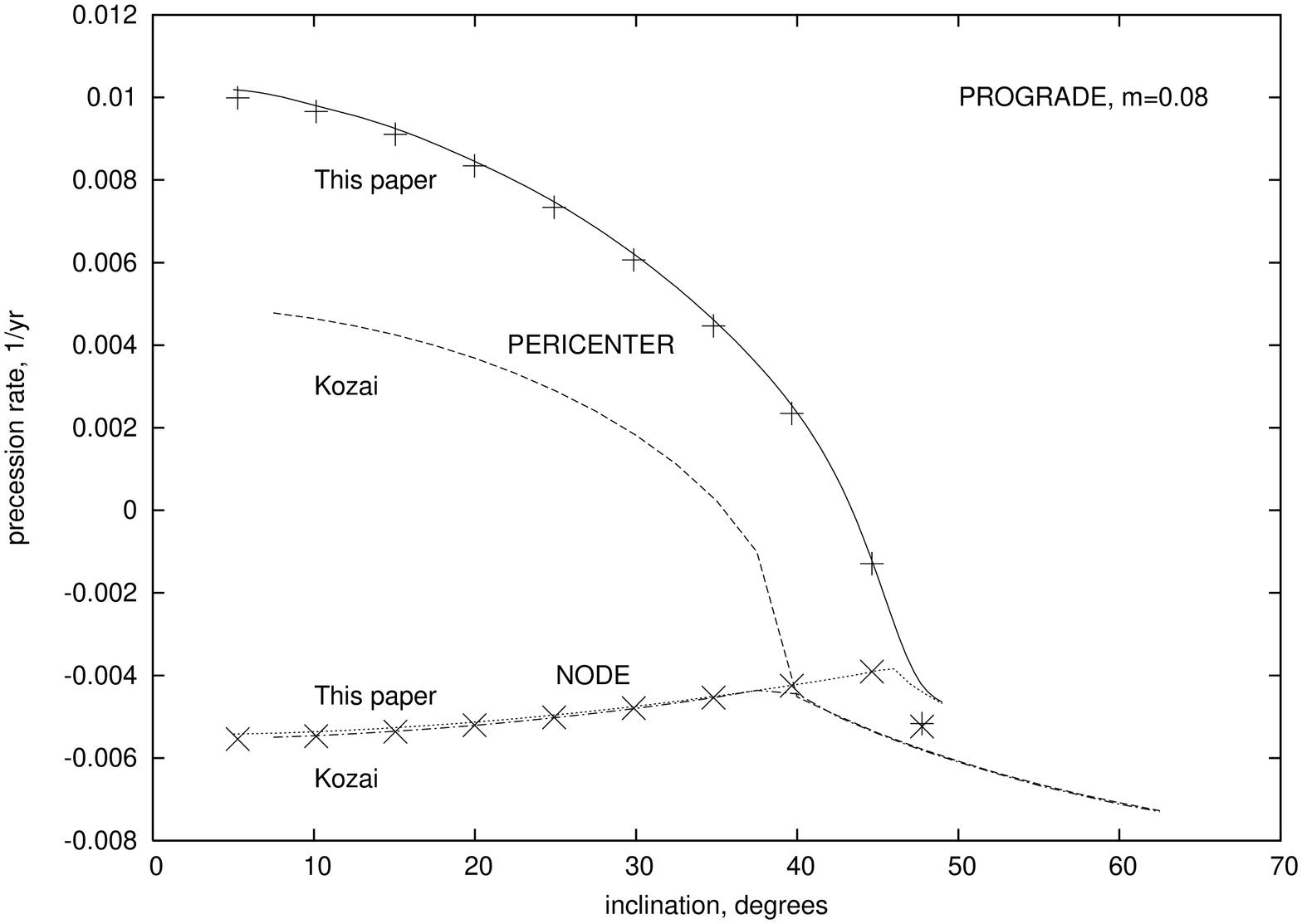}  
\caption{}
\end{figure}
\addtocounter{figure}{-1}
\addtocounter{subfigure}{1}
\begin{figure}
\label{demoiq}
\includegraphics[angle=270, scale=.5]{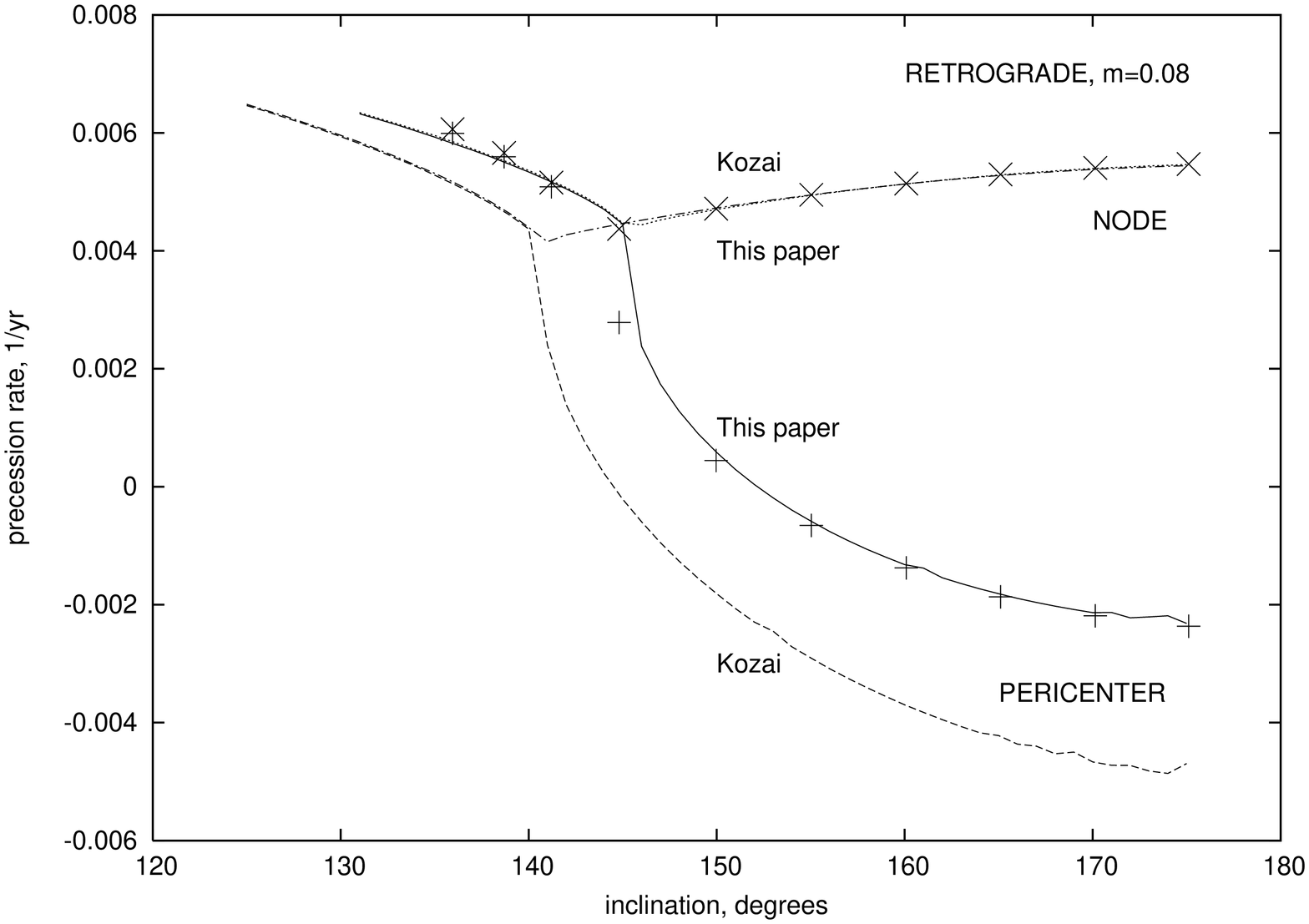}  
\caption{}
\end{figure}
\addtocounter{figure}{-1}
\addtocounter{subfigure}{1}
\begin{figure}
\label{demoir}
\includegraphics[angle=270, scale=.5]{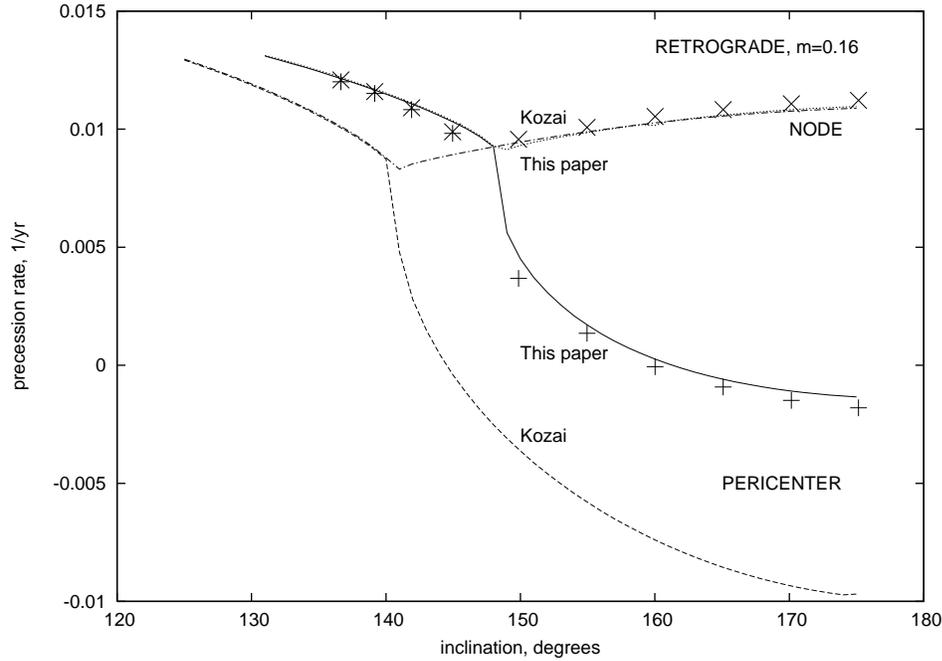}  
\caption{Similar to Fig. 2, but versus inclination, comparing the motion of both the line of apsides and the line of nodes predicted by secular theories with the numerical integrations. The continuous lines and the discrete points have the same meanings as in Figs. 2a-c. All orbits shown are low eccentricity ($e_{max}=0.2$) and have scaled orbital periods of a) $m=0.08$ (prograde), b) $m=0.08$ (retrograde), and c) $m=0.16$ (retrograde). The minor irregularities on some of the curves are artifacts of sampling.}
\end{figure}

\newpage

\renewcommand{\thefigure}{\arabic{figure}\alph{subfigure}}
\setcounter{subfigure}{1}
\begin{figure}
\label{i0_i}
\includegraphics[angle=270, scale=.6]{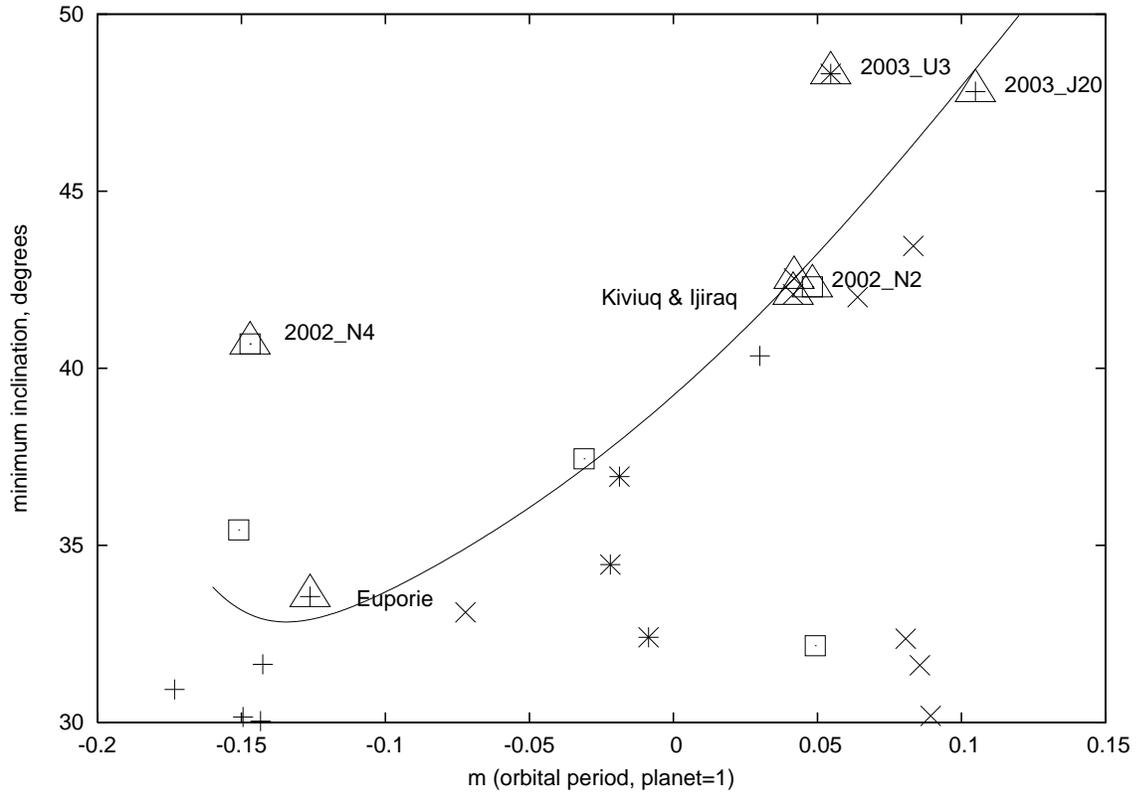}  
\caption{The critical inclination (as a function of $m$) at which Kozai resonance occurs for nearly circular orbits, obtained by twenty iterations of Eq. \ref{iteration} (solid line). The individual points are the known irregular satellites: Jovians (pluses), Saturnians (crosses), Uranians (asterisks) and Neptunians (boxes). The large triangles mark the known Kozai librators, according to numerical simulation, whose names are also labeled.}
\end{figure}
\addtocounter{figure}{-1}
\addtocounter{subfigure}{1}
\begin{figure}
\label{i0_bs}
\includegraphics[angle=270, scale=.6]{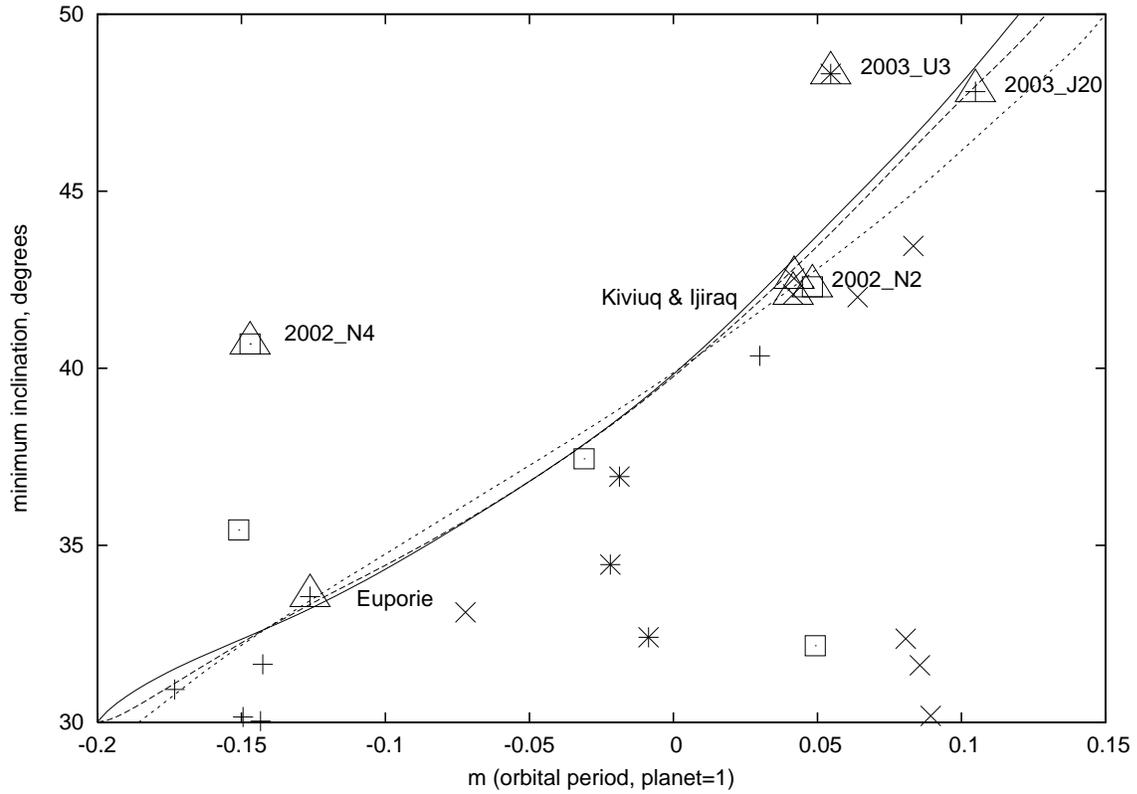}  
\caption{Critical inclination for the Kozai resonance as determined by numerical advancement of the secular equations of motion, for particles with a maximum eccentricity of 0.2 (solid), 0.4 (long dash) and 0.6 (short dash). The symbols for the known individual satellites have the same meaning as in panel a).}
\end{figure}
\renewcommand{\thefigure}{\arabic{figure}}
\begin{figure}
\label{sec_res}
\includegraphics[angle=270, scale=.6]{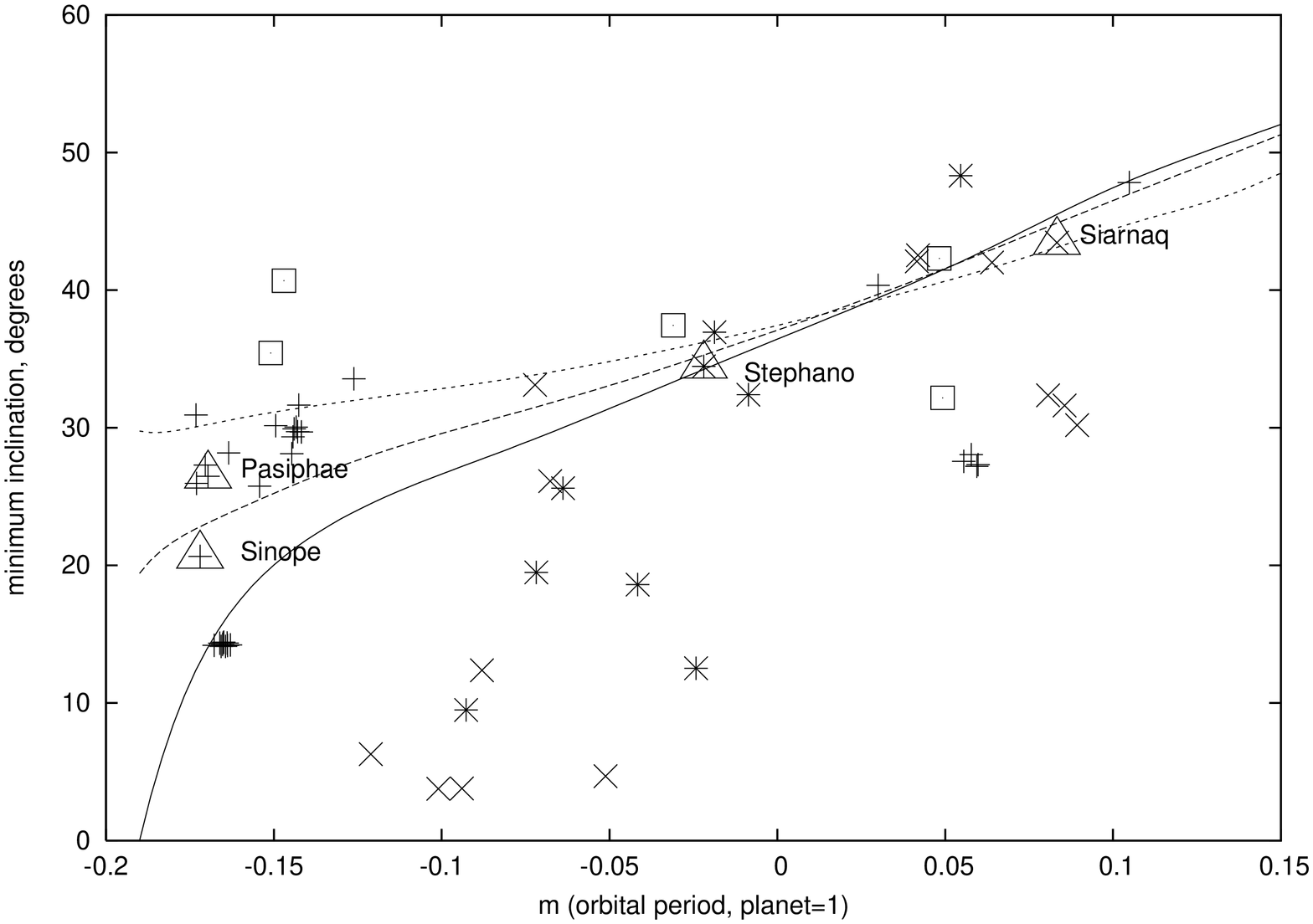}  
\caption{ The location of the secular resonance determined by numerical advancement of the secular equations of motion, for particles with a maximum eccentricity of 0.2 (solid), 0.4 (long dash) and 0.6 (short dash). The individual points are the known irregular satellites: Jovians (pluses), Saturnians (crosses), Uranians (asterisks) and Neptunians (boxes). The large triangles mark the objects known to be in, or close to, secular resonance, with their names also labeled.}
\end{figure}

\clearpage
\renewcommand{\thefigure}{\arabic{figure}\alph{subfigure}}
\setcounter{subfigure}{1}
\begin{figure}
\label{pasiphae}
\includegraphics[angle=270, scale=.5]{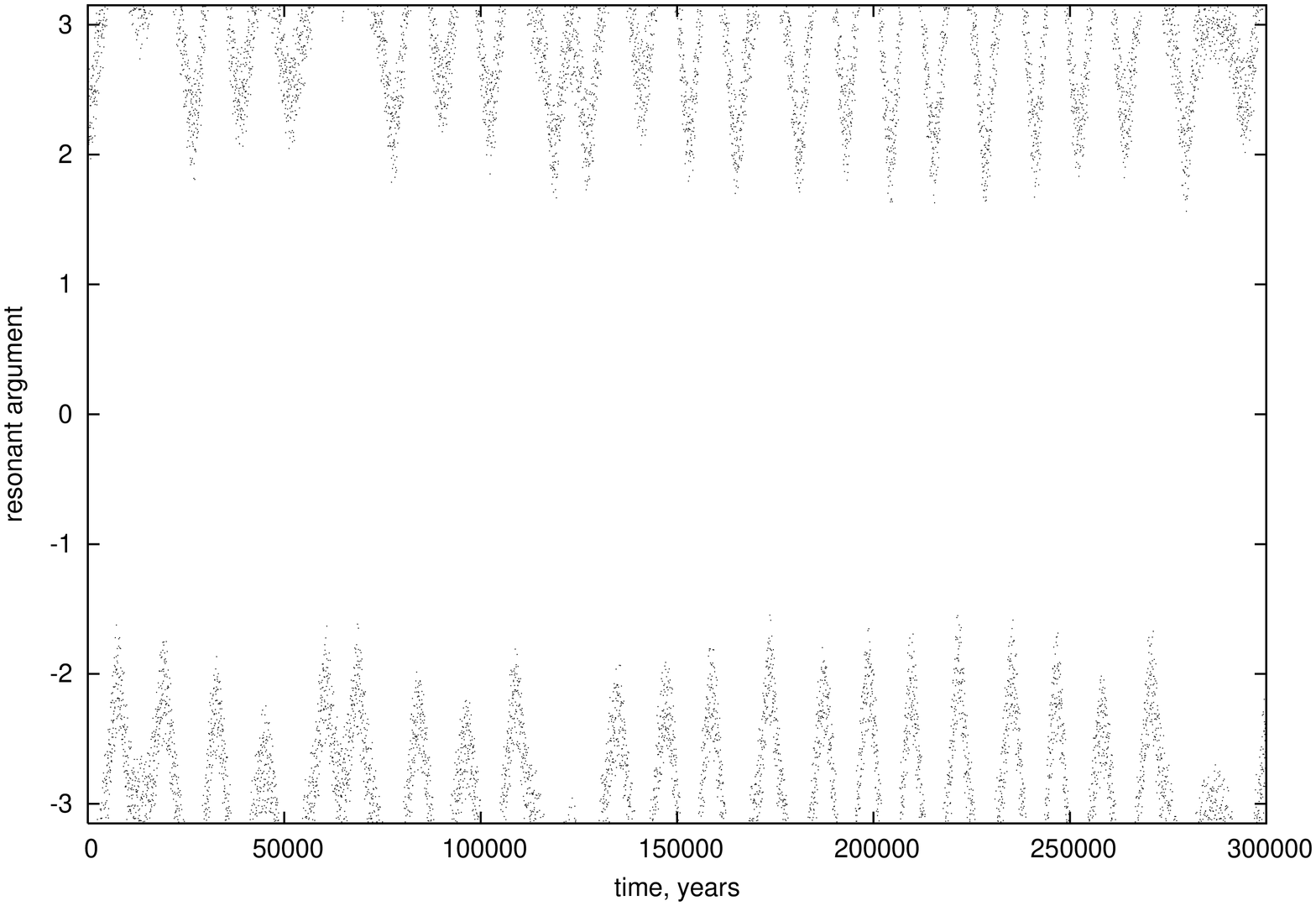}  
\caption{}
\end{figure}

\addtocounter{figure}{-1}
\addtocounter{subfigure}{1}
\begin{figure}
\label{sinope}
\includegraphics[angle=270, scale=.5]{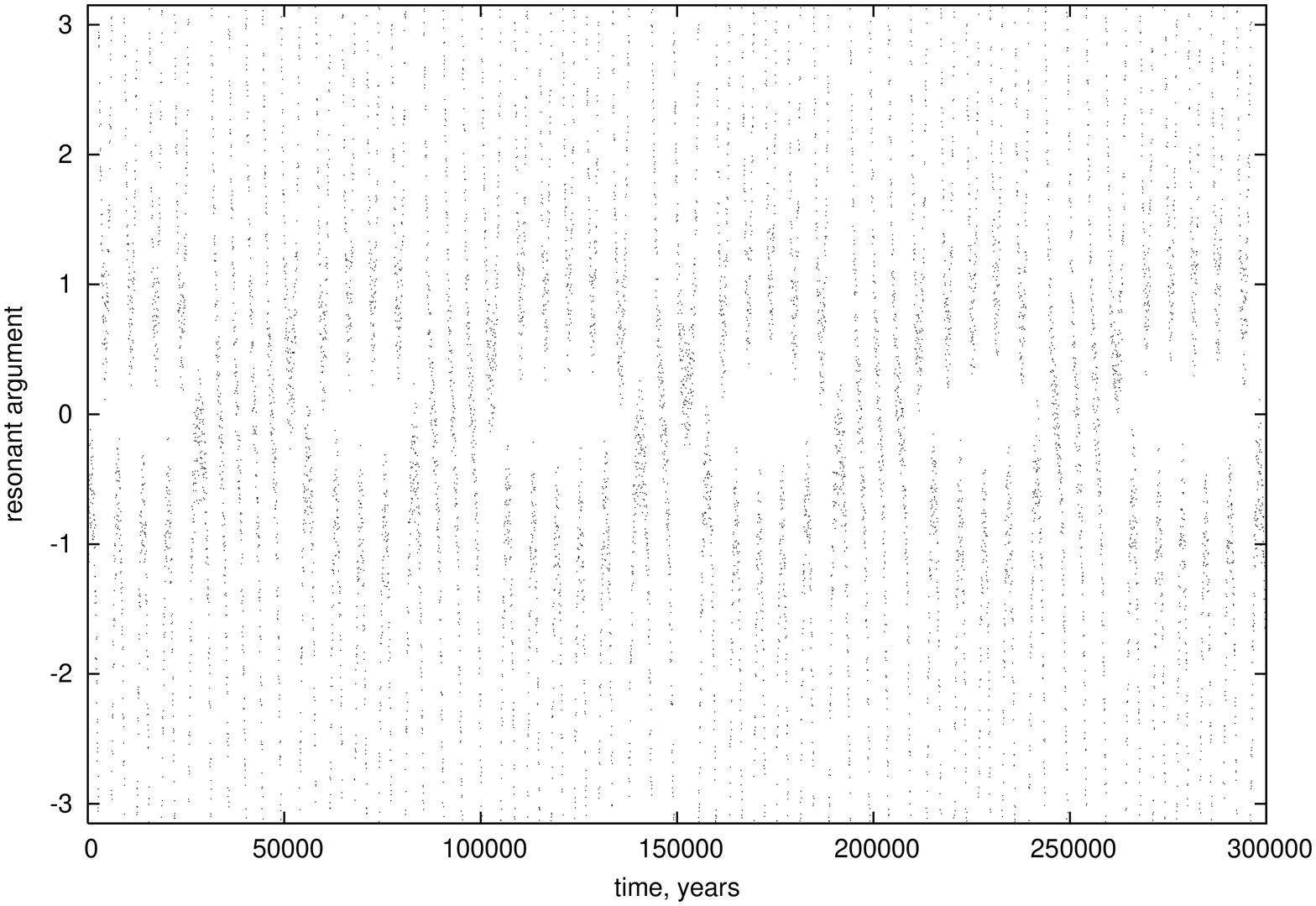}  
\caption{}
\end{figure}

\addtocounter{figure}{-1}
\addtocounter{subfigure}{1}
\begin{figure}
\label{siarnaq}
\includegraphics[angle=270, scale=.5]{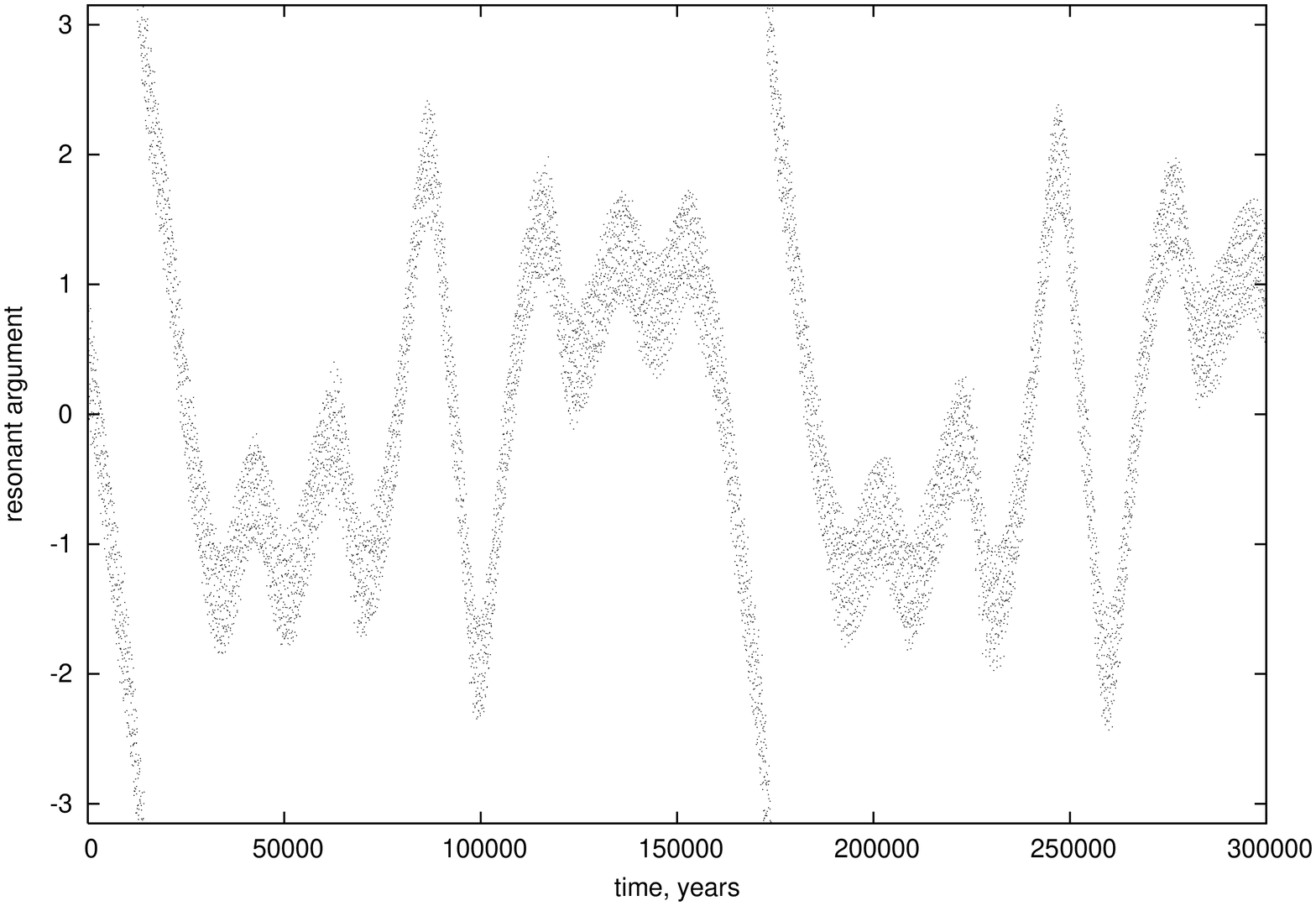}  
\caption{}
\end{figure}

\addtocounter{figure}{-1}
\addtocounter{subfigure}{1}
\begin{figure}
\label{stephano}
\includegraphics[angle=270, scale=.5]{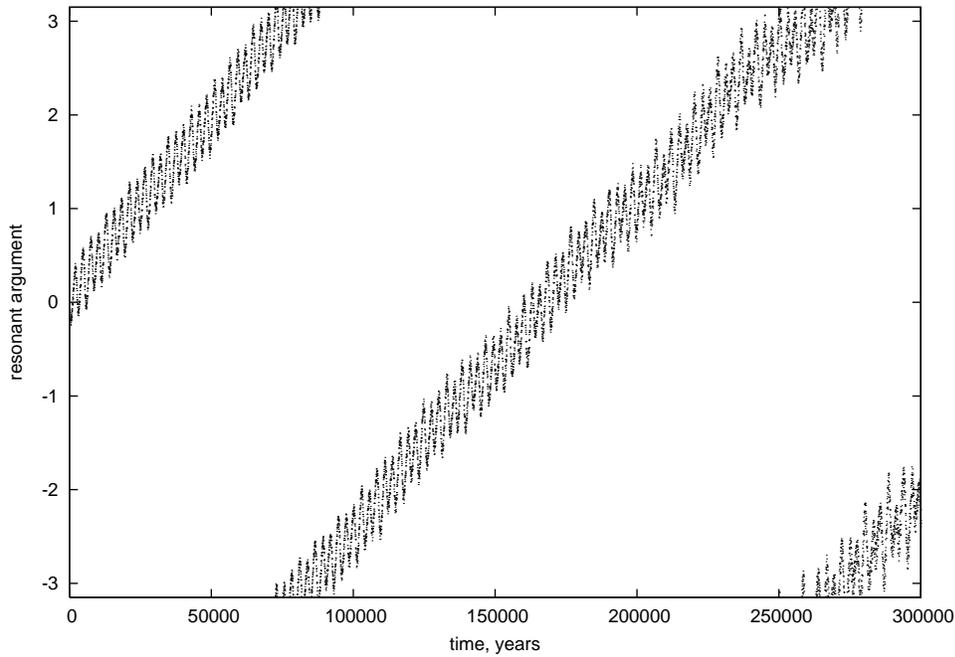}  
\caption{Evolution of the resonant argument $\Psi=\varpi-\varpi_{planet}$ (in radians) during a 300,000-yr numerical integration for four resonant or near-resonant satellites: a) Jupiter's Pasiphae, b) Jupiter's Sinope, c) Saturn's Siarnaq and d) Uranus's Stephano.}
\end{figure}

\clearpage
\renewcommand{\thefigure}{\arabic{figure}\alph{subfigure}}
\setcounter{subfigure}{1}
\begin{figure}
\label{sinope_e}
\includegraphics[angle=270, scale=.49]{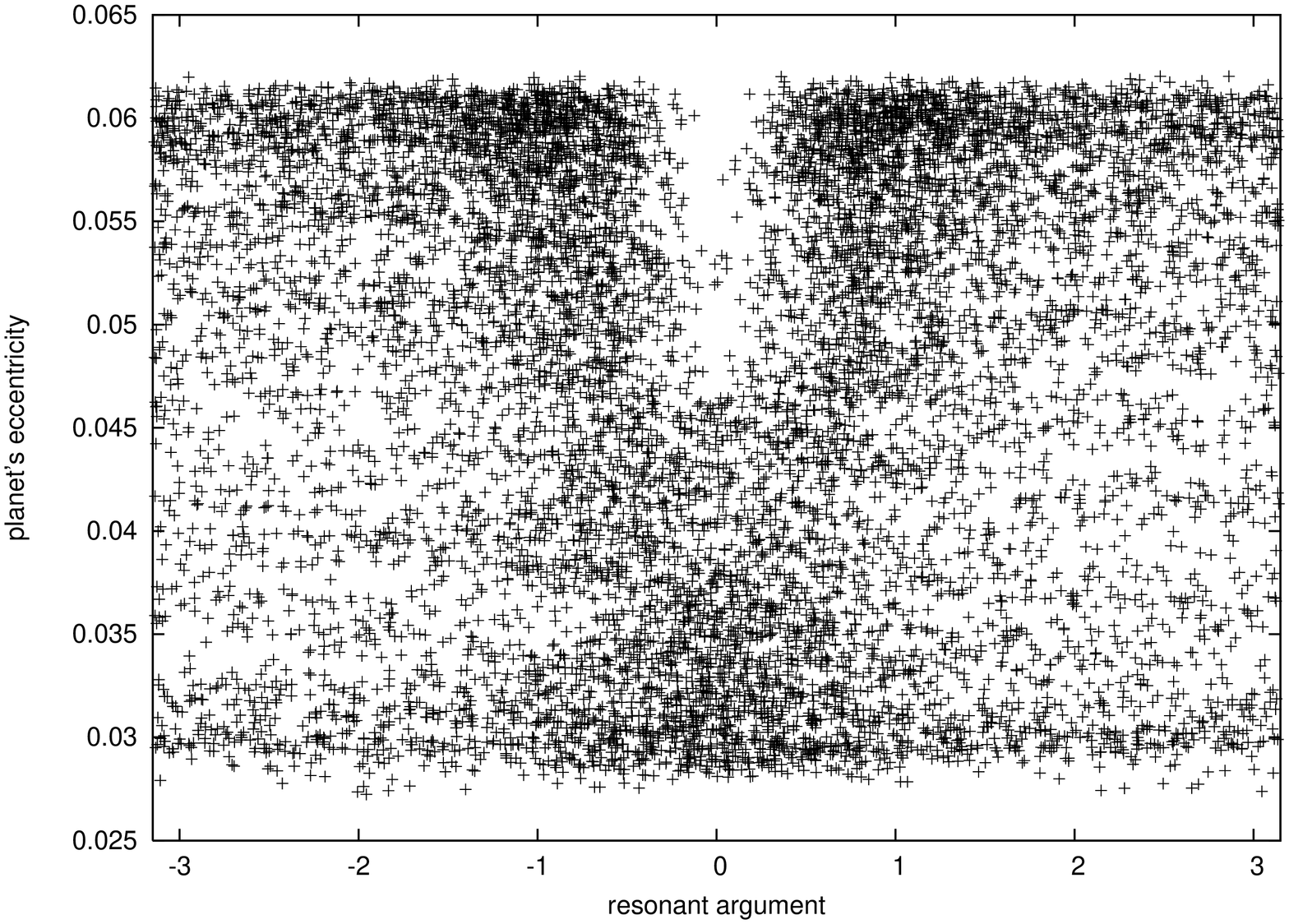}
\caption{}
\end{figure}

\addtocounter{figure}{-1}
\addtocounter{subfigure}{1}
\begin{figure}
\label{siarnaq_e}
\includegraphics[angle=270, scale=.49]{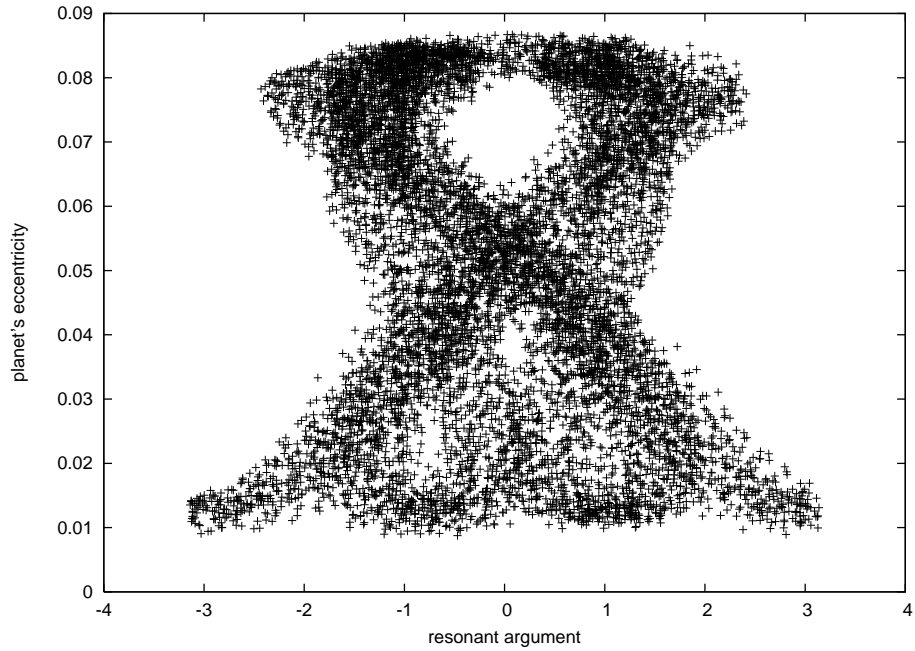}  
\caption{Correlation between the resonant argument $\Psi$ and the parent planet's eccentricity for a) Jupiter's Sinope and b) Saturn's Siarnaq. Note that Sinope's $\Psi$ avoids passing through $\Psi=0$ when Jupiter's eccentricity is in its upper range, whereas Siarnaq's $\Psi$ passes through $\Psi=\pi$ only when Saturn's eccentricity is at minimum.}
\end{figure}
\clearpage

\renewcommand{\thefigure}{\arabic{figure}}
\begin{figure}
\label{Scosi}
\includegraphics[angle=270, scale=.6]{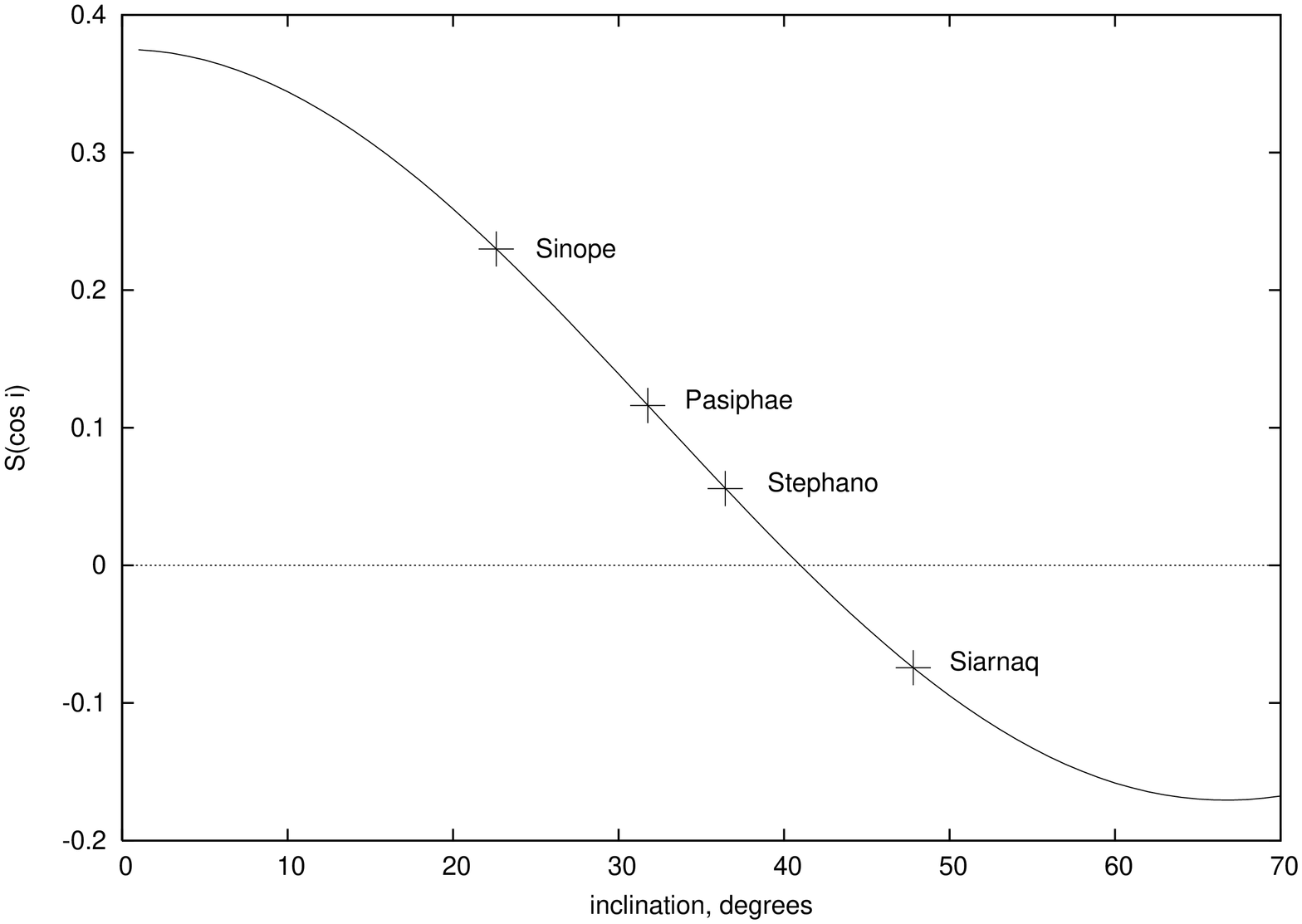}
\caption{The dependence of $S (\cos i)$ as a function of the orbital inclination (defined here as being between $0^\circ$ and $90^\circ$). The values of $S(\cos i)$ and $i$ for four resonant and near-resonant objects are also labeled. $S (\cos i)$ passes through zero at an average inclination of about $41^\circ$ (139$^\circ$ for retrograde orbits).}
\end{figure}
\renewcommand{\thefigure}{\arabic{figure}\alph{subfigure}}
\setcounter{subfigure}{1}
\clearpage
\begin{figure}
\label{gi_e}
\includegraphics[angle=270, scale=.50]{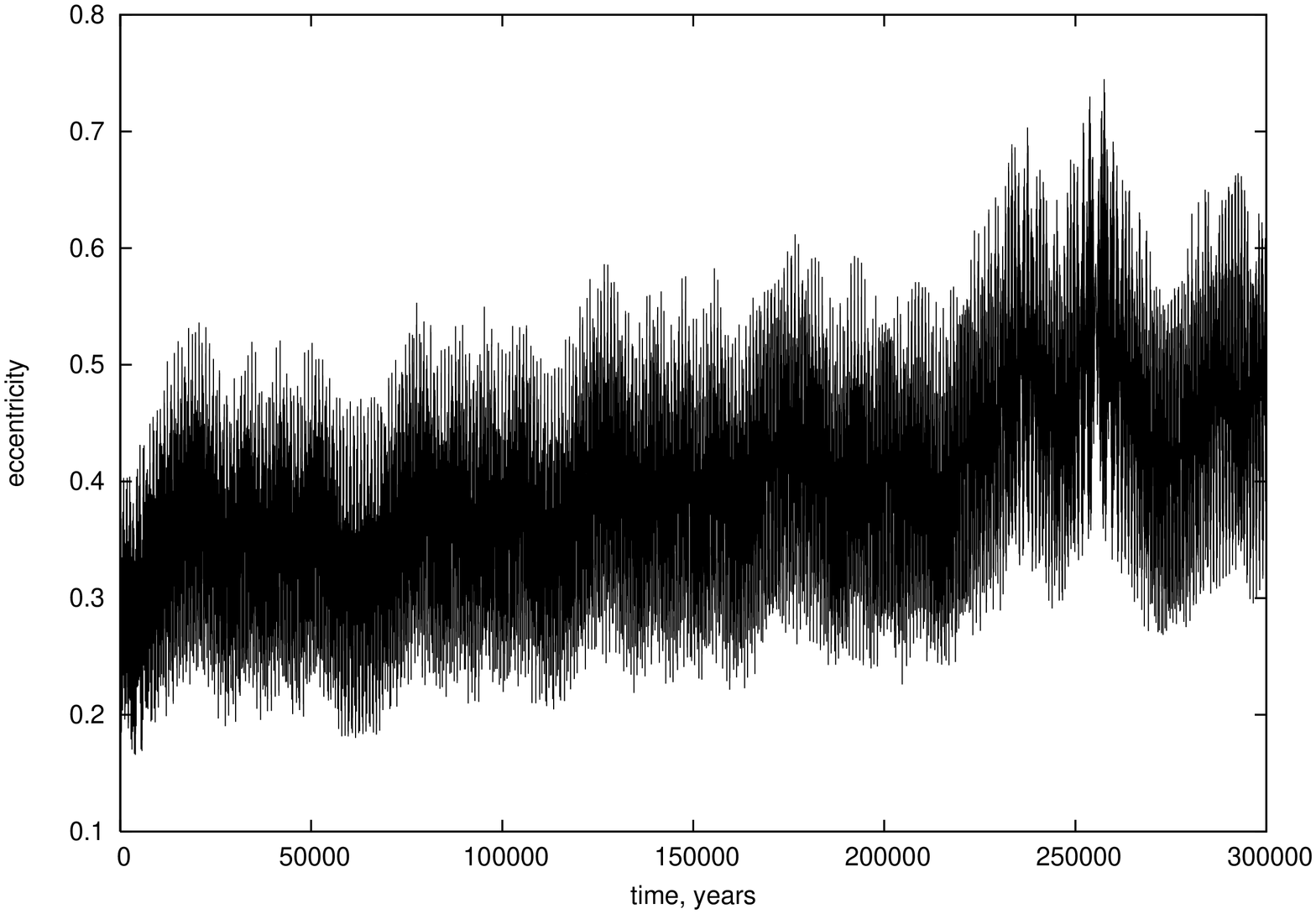}
\caption{}
\end{figure}
\addtocounter{figure}{-1}
\addtocounter{subfigure}{1}
\begin{figure}
\label{gi_xi}
\includegraphics[angle=270, scale=.50]{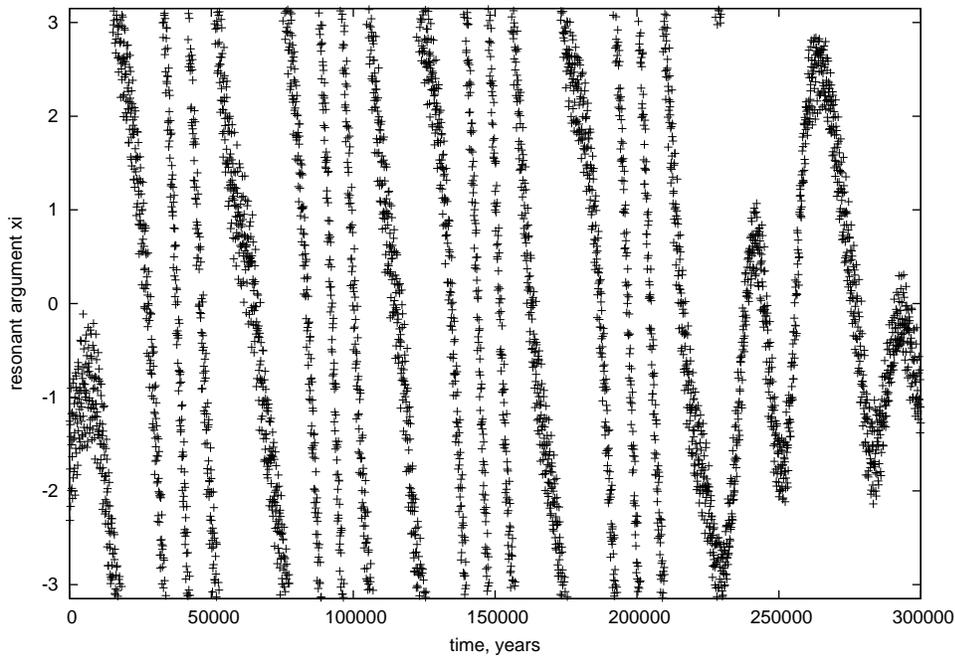}  
\caption{a) Evolution of eccentricity over a 300,000-yr numerical integration for a particle orbiting Saturn whose orbit is affected by the Great-Inequality resonance. b) Evolution of the resonant argument $\xi=5 \lambda_S - 2 \lambda_J - 2 \varpi - \varpi_S$ during the same integration.
}
\end{figure}

\clearpage
\renewcommand{\thefigure}{\arabic{figure}}
\begin{figure}
\label{gi_eta}
\includegraphics[angle=270, scale=.6]{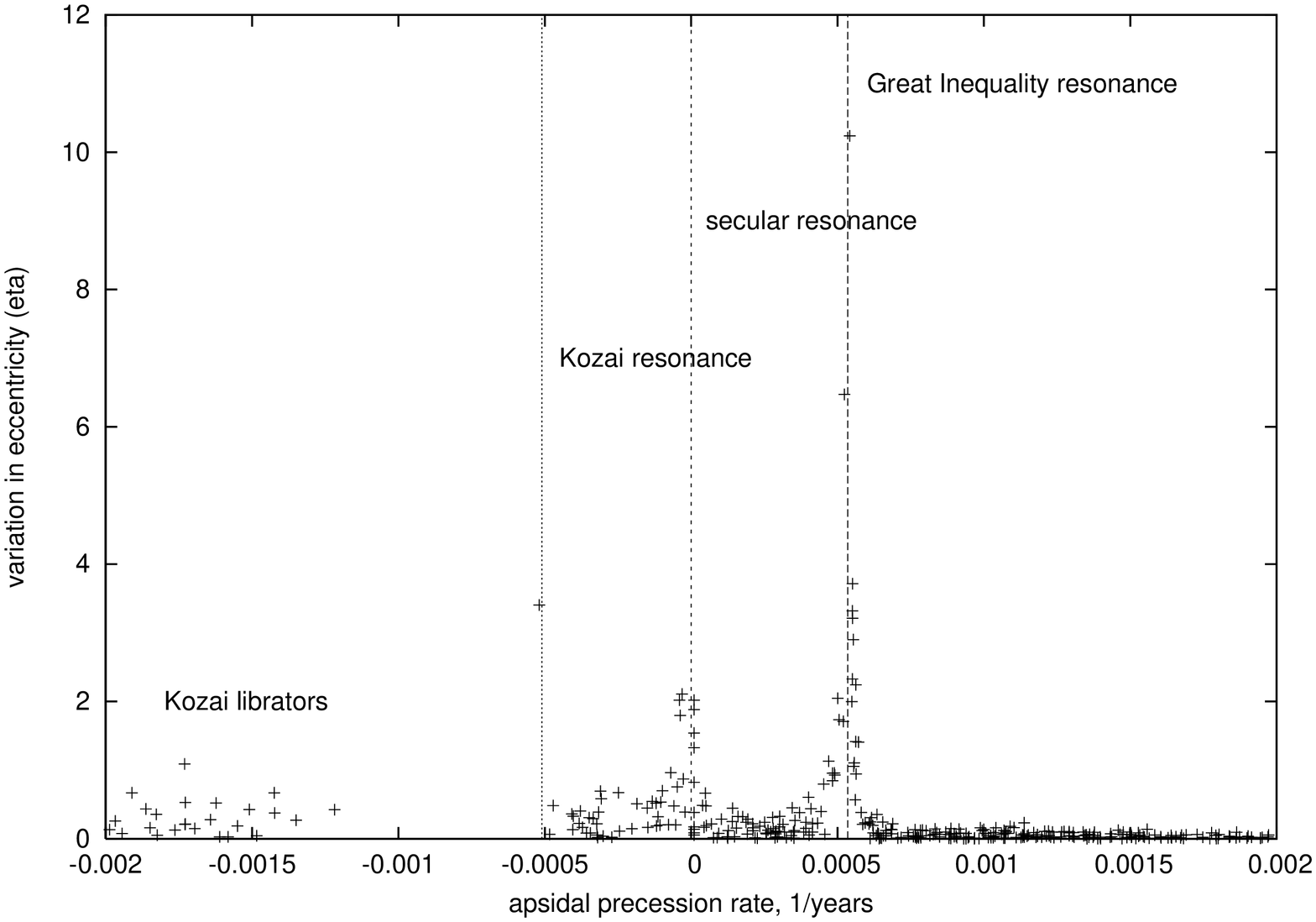}
\caption{The dependence of eccentricity variation parameter $\eta$ (defined in Eq. \ref{eta}) on the precession rate of a particle's $\varpi$ for 440 test particles integrated numerically over 30,000 yrs (see text for details). Secular and the Great Inequality resonances are prominent as regions of high $\eta$; the onset of Kozai resonance is also visible, with all the particles to the left of the discontinuity being Kozai librators.}
\end{figure}
\clearpage
\begin{figure}
\label{gi_mi}
\includegraphics[angle=270, scale=.6]{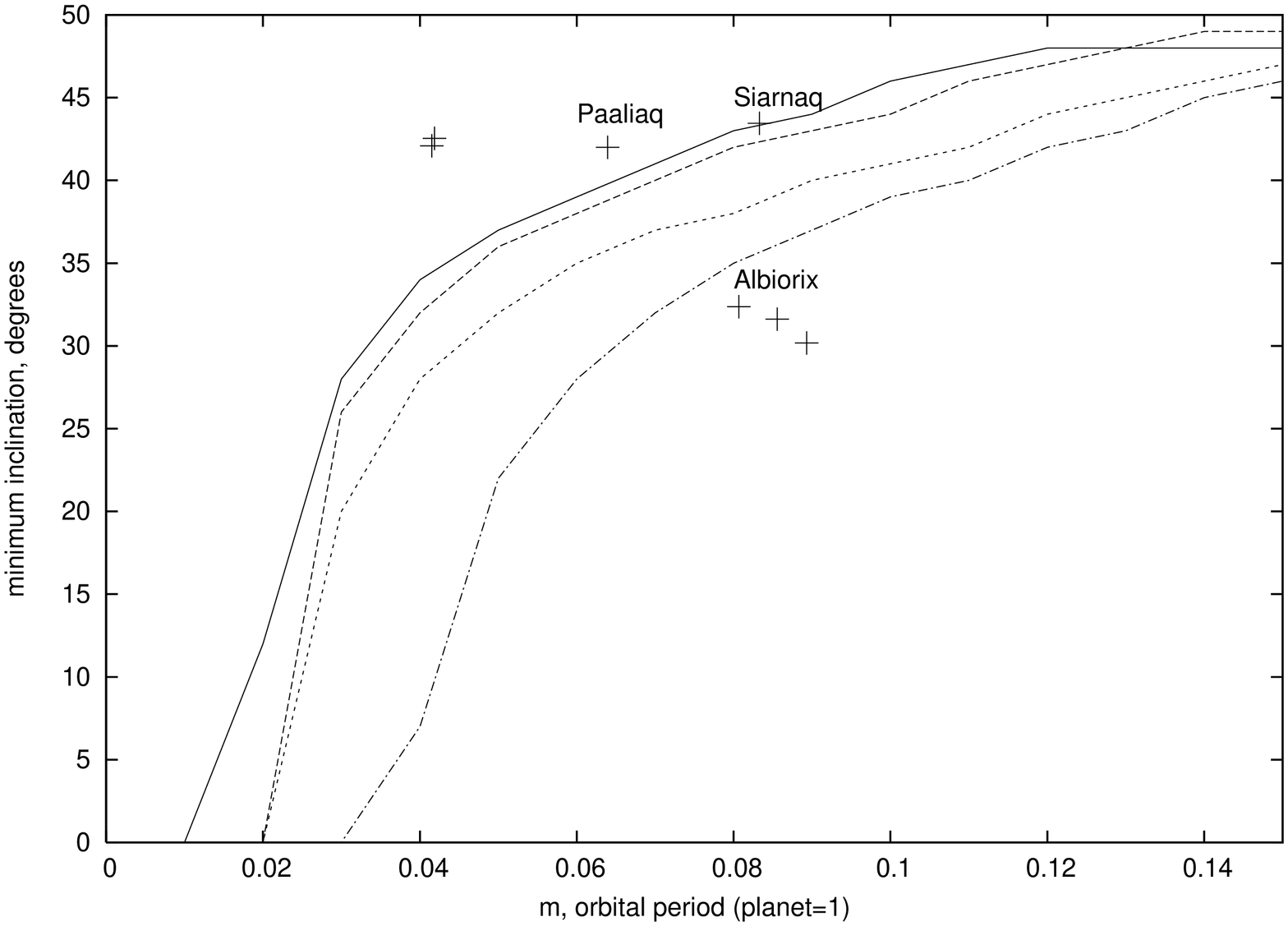}
\caption{The position of the Great-Inequality resonance in $m-i_{min}$ space. The top three continuous lines plot (in descending order) the Great Inequality resonance for maximum eccentricities of 0.2, 0.4 and 0.6, respectively. The fourth line plots the position of the Great Inequality resonance for $e_{max}=0.6$, in a planetary configuration that has $5 \lambda_S - 2 \lambda_J=0.002$ yr$^{-1}$ (i.e., the Great Inequality has a period of 500 yrs). The orbital elements of the known prograde Saturnian satellites are also shown as individual points.}
\end{figure}
\begin{figure}
\label{MS}
\includegraphics[angle=270, scale=.6]{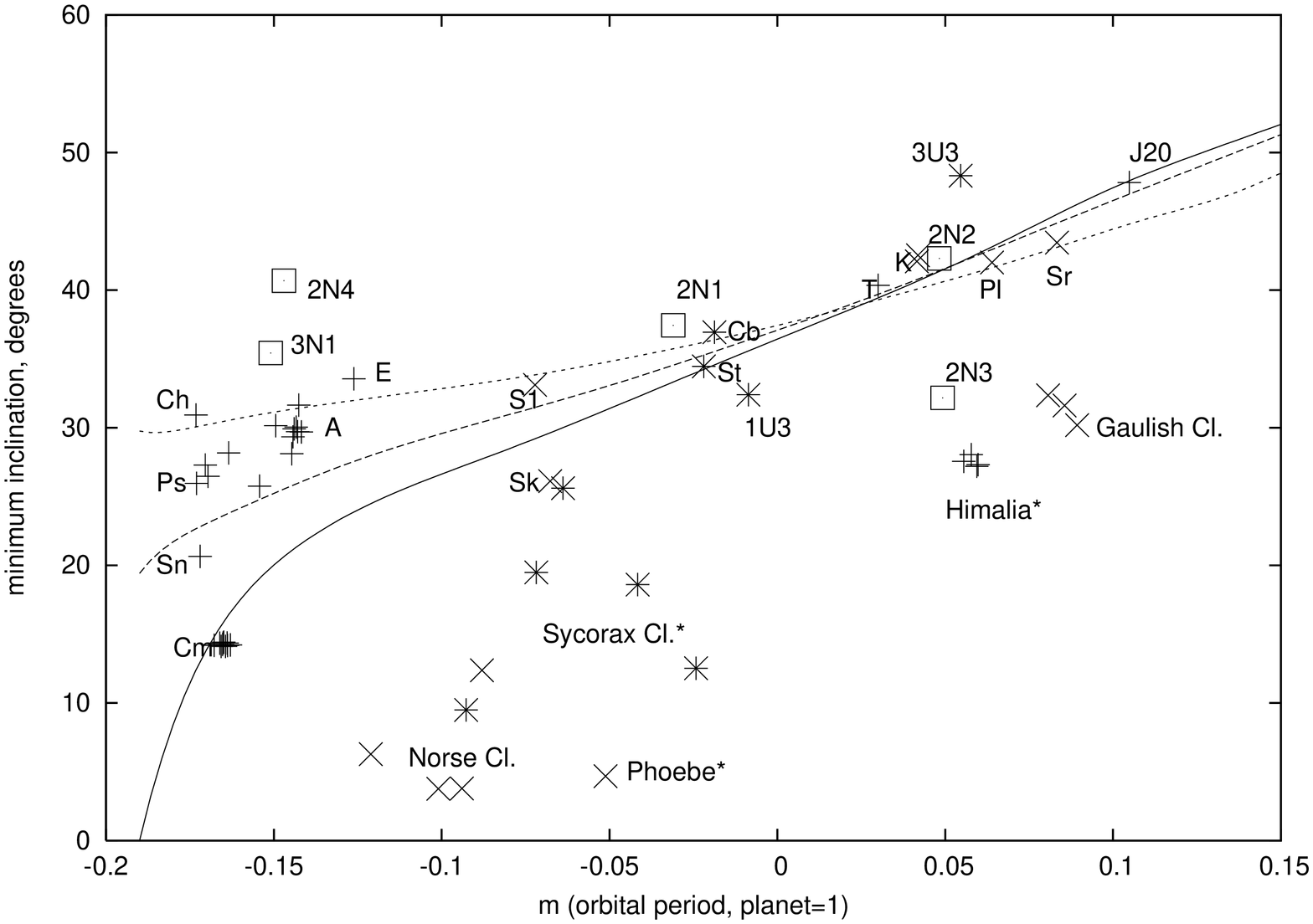}
\caption{Same as Fig. 5, only with all satellite clusters and isolated satellites labeled. The groups whose names are followed by an asterisk contain objects with $R > 100$ km. The clustering of groups around the secular resonance is clearly visible. See text for the full explanation of the abbrevations used in this figure. }
\end{figure}
\renewcommand{\thefigure}{\arabic{figure}\alph{subfigure}}
\setcounter{subfigure}{1}
\begin{figure}
\label{sinope_hk}
\includegraphics[angle=270, scale=.45]{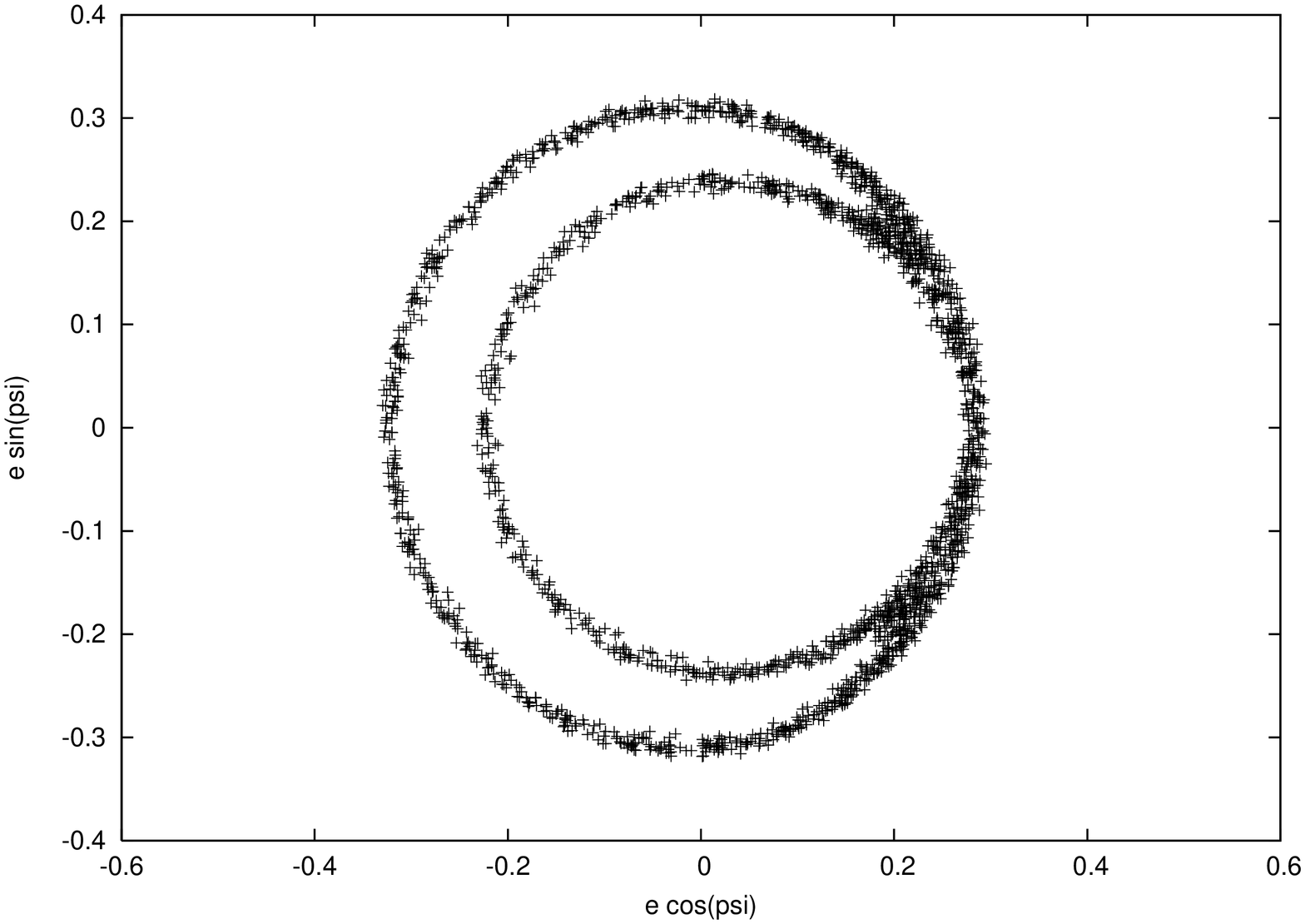}
\caption{}
\end{figure}

\addtocounter{figure}{-1}
\addtocounter{subfigure}{1}
\begin{figure}
\label{sinope_e100}
\includegraphics[angle=270, scale=.45]{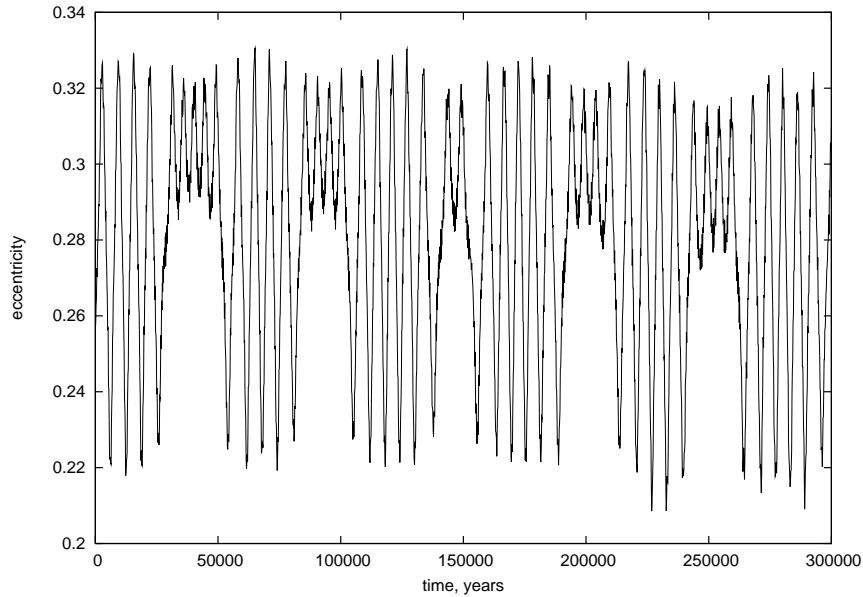}  
\caption{a) The evolution of Sinope's $e$ and its resonant argument $\Psi=\varpi-\varpi_J$ during the numerical integration shown in Fig. 6b, plotted in $h=e \cos \Psi$ and $k=e \sin \Psi$ coordinates. The two rings represent two modes of circulation that are characteristic of the regions on either side of the secular resonance, while the librations show as crescent-shaped figures that include portions of each ring. b) Evolution of $e$ over time in the same simulation. Note that the episodes of high-$e$ in this plot are correlated with the instances of retrograde circulation of $\Psi$ in Fig 6b.}
\end{figure}

\clearpage 

\begin{table}
	\caption{The strength of the resonant term for irregular satellites known to be in, or close to, secular resonance.}
	\label{resonators}
        \begin{tabular}{|l|c|c|c|c|c|c|}
            \tableline\tableline
            Satellite & $a/a'$ & $<e>$ & $b_1(e)$ & $<i> [^\circ]$ & $S(\cos i)$ & $R'_r/R'_{r, P}$\\
            \tableline
            Pasiphae & 0.03014 & 0.4106 &-1.1563 &148.24 & 0.11617 & 1.0000\\
            Sinope & 0.03041 & 0.2891 &-0.7681 &157.39 & 0.22990 & 1.3262\\
            Siarnaq & 0.01251 & 0.3041 &-0.8130 & 47.81 &-0.07441 &-0.1870 \\
            Stephano & 0.00277 & 0.2320 &-0.6033 &143.55 & 0.05582 & 0.0230\\
            \tableline
        \end{tabular}
\end{table}

\begin{table}
	\caption{The ratio $\nu=-\dot{\varpi}/\dot{\Omega}$ and our dynamical classification for some irregular satellites. Categories are secular resonators (SR), Kozai librators (K), reverse-circulators (RC), ``Main Sequence'' circulators (MSC) and non-''Main Sequence'' objects (non-MS). $\nu$ was computed based on our direct numerical integrations described in Sec. 3, except for S/2003 S1, where an updated orbital integration was based on \citet{she04}}
	\label{criterion}
        \begin{tabular}{|l|c|c||l|c|c|}
            \tableline\tableline
            Satellite & $\nu$ & Class & Satellite & $\nu$ & Class\\
            \tableline
 Himalia &	 1.10 &	non-MS	&  Kiviuq &	-1.00 &	K\\
 Pasiphae &	 0.00 &	SR	&  Ymir	&	 0.28 &	non-MS\\
 Sinope	&	 0.00 &	SR	&  Skadi &	 0.16 &	MSC\\
 Carme	&	 0.09 &	MSC	&  S/2003 S1 &	 0.00* & SR (?)\\
 Ananke	&	-0.24 &	RC	&  Sycorax &     0.35 &	non-MS\\
 Callirrhoe &	-0.26 &	RC	&  Caliban &	-0.26 &	RC\\
 Themisto &	-0.21 &	RC	&  S/2001 U3 &	 0.19 &	MSC\\
 Euporie &	-1.00 &	K	&  S/2003 U3 &	-1.00 &	K\\
 S/2003 J20 &	-1.00 &	K	&  S/2002 N1 &	-0.04 &	RC (?)\\
 Phoebe	&    	 0.59 &	non-MS	&  S/2002 N2 &	-1.00 &	K\\
 Siarnaq &	 0.00 & SR	&  S/2002 N3 &	 0.33 &	non-MS\\
 Albiorix &	 0.41 &	non-MS	&  S/2002 N4 &	-1.00 &	K\\
 Paaliaq &	-0.08 &	RC	&  S/2003 N1 &  -0.10 & RC\\
            \tableline
        \end{tabular}
\end{table}

\end{document}